\documentclass[apj,twocolumn,revtex4,twocolappendix]{emulateapj} 
\usepackage{graphicx}
\usepackage{bm}
\usepackage[dvips]{color}
\usepackage{xspace}
\usepackage{ulem}
\usepackage{lscape}

\renewcommand\thefootnote{\alpha{footnote}}

\newcommand{\ironc}{$L_{\rm K\alpha}/L_{\rm 10-50\hspace{0.05cm}keV}$}

\shorttitle{{\it SUZAKU} OBSERVATIONS OF LOW-LUMINOSITY AGNS}
\shortauthors{Kawamuro et al.}

\begin{document}
\title{STUDY OF SWIFT/BAT SELECTED LOW-LUMINOSITY ACTIVE GALACTIC NUCLEI OBSERVED WITH SUZAKU} 

\author{
 Taiki Kawamuro\altaffilmark{1},
 Yoshihiro Ueda\altaffilmark{1},
 Fumie Tazaki\altaffilmark{2},
 Yuichi Terashima\altaffilmark{3}, 
 Richard Mushotzky\altaffilmark{4},
}

\altaffiltext{1}{Department of Astronomy, Kyoto University, Kyoto 606-8502, Japan}
\altaffiltext{2}{Mizusawa VLBI Observatory, National Astronomical Observatory of Japan, 
Osawa, Mitaka, Tokyo 181-8588, Japan}
\altaffiltext{3}{Department of Physics, Ehime University, Matsuyama 790-8577, Japan}
\altaffiltext{4}{Department of Astronomy, University of Maryland, CollegePark, MD 20742-2421, USA}

\begin{abstract} 

We systematically analyze the broadband (0.5--200 keV) X-ray spectra of
hard X-ray ($>10$ keV) selected local low-luminosity active galactic
nuclei (LLAGNs) observed with {\it Suzaku} and {\it Swift}/BAT. The
sample consists of ten LLAGNs detected with {\it Swift}/BAT with
intrinsic 14--195 keV luminosities smaller than $10^{42}$ erg s$^{-1}$
available in the {\it Suzaku} archive, covering a wide range of the
Eddington ratio from $10^{-5}$ to $10^{-2}$. The overall spectra can be
reproduced with an absorbed cut-off power law, often
accompanied by reflection components from distant cold matter, and/or
optically-thin thermal emission from the host galaxy. In all objects,
relativistic reflection components from the innermost disk are not
required. Eight objects show a significant narrow iron-K$\alpha$
emission line. Comparing their observed equivalent widths with the
predictions from the Monte-Carlo based torus model by \cite{Ike09}, we
constrain the column density in the equatorial plane to be $\log N^{\rm
eq}_{\rm H} > 22.7$ or the torus half opening angle $\theta_{\rm oa} <
70^\circ$. We infer that the Eddington ratio ($\lambda_{\rm Edd}$) 
is a key parameter that determines the torus structure of LLAGNs: the torus
becomes large at $\lambda_{\rm Edd} \gtrsim 2\times10^{-4}$, whereas at
lower accretion rates it is little developed. The luminosity correlation
between the hard X-ray and mid-infrared (MIR) bands of the LLAGNs
follows the same one as for more luminous AGNs. This implies that other
mechanisms than AGN-heated dust are responsible for the MIR emission in
low Eddington ratio LLAGNs.

\end{abstract}


\section{INTRODUCTION}

Active galactic nuclei (AGNs) emit intense X-ray radiation by converting the 
gravitational energy of matter accreted onto the supermassive black holes 
(SMBHs). According to the so-called AGN unified model \citep{Ant93}, the 
central engine is obscured by a dusty torus, which affects the observed X-ray 
spectrum via photoelectric absorption and Compton reflection. In addition, a 
relativistically blurred reflection component from the innermost accretion 
disk would be expected when the disk is extended down to a vicinity of the 
SMBH. Hence, observations of broadband X-ray spectra of AGNs are useful to 
investigate their structure including the torus and accretion disk.

Low-luminosity AGNs (LLAGNs), which we define by their low X-ray luminosities 
in the 14--195 keV band ($L_{\rm X} < 10^{42}$ erg s$^{-1}$), are important 
objects to understand the 
evolution of AGNs. LLAGNs contain two extreme types of AGNs: those with a 
small SMBH mass and a high mass-accretion rate (i.e., with a high Eddington 
ratio), and those with a large SMBH mass and a low mass-accretion rate (a low 
Eddington ratio). The former type is expected in the early growing phase of a 
SMBH, whereas the latter corresponds to the fading phase of an AGN into a 
quiescent SMBH. Broadband X-ray studies of LLAGNs have been limited, however, 
because of their low fluxes even in the local Universe (\citealt{Kawa13}). 

Theories and observations suggest that LLAGNs are not a simple scaled-down 
version of luminous AGNs in terms of their nucleus structure. When the 
accretion rate falls below a critical level, it is predicted that the state 
of the inner accretion disk changes from the standard accretion disk 
\citep{Sha73} to an optically-thin radiatively inefficient accretion flow (RIAF; 
\citealt{Nara98}; \citealt{Qua01}). Studies of spectral energy distributions (SEDs) from the 
radio to X-ray bands showed that the standard disk is truncated at a radius 
much larger than a few Schwarzschild radii (e.g., \citealt{Nem11}; 
\citealt{Mas12}). This is supported by the lack of the big blue bump in LLAGNs, 
which arises from the thermal emission from the inner standard disk (e.g.,
\citealt{Ho08}). Moreover, most of LLAGNs do not show a broad iron-K$\alpha$ 
line feature, suggesting that the standard disk does not extend to the inner 
most region around the SMBH, except for some LLAGNs (e.g., NGC 4051; 
\citealt{Gua96}). Also, \cite{Gu09} found a negative correlation between the 
X-ray photon index and Eddington ratio for low Eddington ratio AGNs 
($\lambda_{\rm Edd} < 0.01$), whereas the positive one for luminous AGNs was 
reported by \cite{She08}. \cite{Gu09} suggested that the negative correlation 
can be explained with a RIAF model. For the detailed investigations of the 
broad iron line feature and continuum components, broadband X-ray spectra are 
highly important. 

X-ray surveys showed that the absorbed AGN fraction, which reflects the torus 
covering factor, decreases with X-ray luminosity (e.g., \citealt{Ued03}, 2014; 
\citealt{Del08}; \citealt{Has08}; \citealt{Mer14}; \citealt{Buc15}).
These results would be consistent with receding torus models
 (e.g., \citealt{Law91}), where the inner wall of the torus recedes with 
luminosity by keeping its scale-height more constant. However, recent surveys 
have found evidence that the absorbed AGN fraction decreases towards lower 
X-ray luminosities from a peak around $L_{\rm X} \sim 10^{42-43}$ erg s$^{-1}$ 
(e.g., \citealt{Bec09}; \citealt{Bur11}; \citealt{Bri11b}). These facts 
suggest that, at low luminosities, other physical mechanisms than AGN 
radiation must be responsible to determine the torus structure. 
 
Main goals of this paper are (1) to best constrain the X-ray
spectral properties of LLAGNs by taking advantage of their ``broadband''
(0.5--200 keV) data observed with {\it Suzaku} and {\it Swift}/BAT, (2) to reveal
the nuclear structure including the torus and accretion disk, and (3) to
find key parameters that determine them. Most of previous studies are limited to soft energy 
bands below $\sim$10 keV (e.g., \citealt{Gon09}).
From the {\it Swift}/BAT 70-month catalog \citep{Bau13}, we select all ten 
LLAGNs ($\log L_{\rm X} < 42$ in the 14--195 keV band) whose {\it Suzaku} 
archival data are public. We exclude NGC~4051, which is known to exhibit large 
spectral variation among different epochs (e.g., \citealt{Gua96}; 
\citealt{Pou04}; \citealt{Tera09}). This paper is organized as follows. 
Section \ref{sec:obs} describes the details of the observation and the 
overview of the data. We show the results of the spectral analysis in Section 
\ref{sec:spec_ana}. The discussion and conclusions are presented in 
Sections~\ref{sec:dis_con} and \ref{sec:con}, respectively. 
Unless otherwise noted, all errors are quoted at the 90\% confidence level for
 a single parameter of interest.


\begin{deluxetable*}{ccccccccc}
\tabletypesize{\small}
\tablecaption{Basic Information of the Targets\label{tab:info_srcs}}
\tablewidth{0pt}
\tablehead{ Target Name  & SWIFT ID & Type & R.A.(J2000) & 
Decl. (J2000) & Redshift  & $D$  & 
$\log M_{\rm BH}/M_{\rm sol}$ & Ref.  \\ 
(1)  & (2) & (3) & (4) & (5) & (6) & (7)  & (8) & (9)
}
\startdata
NGC~2655 & J0856.0+7812 & Seyfert 2   & 08h55m37.7s  & +78d13m23s & 4.670$\times10^{-3}$    & 24.4  & 7.7$^{+0.1}_{-0.2}$ & (1,1) \\
NGC~3718 & J1132.7+5301 & XBONG       &  11h32m34.9s  & +53d04m05s & 3.312$\times10^{-3}$   & 17.0  & 8.0$\pm0.3$       & (1,2) \\
NGC~3998 & J1157.8+5529 & Seyfert 1/LINER & 11h57m56.1s & +55d27m13s & 3.496$\times10^{-3}$ & 19.4  & 8.9$\pm0.1$ & (2,3)  \\
NGC~4102 & J1206.2+5243 &  LINER      & 12h06m23.0s & +52d42m40s  & 2.823$\times10^{-3}$    & 19.1  & 7.9$^{+0.1}_{-0.2}$ & (3,1) \\ 
NGC~4138 & J1209.4+4340 & Seyfert 1.9 & 12h09m29.8s & +43d41m07s & 2.962$\times10^{-3}$     & 15.9  & 7.2$^{+0.1}_{-0.2}$ & (4,1) \\
NGC~4258 & J1219.4+4720 & Seyfert 1.9/LINER & 12h18m57.5s & +47d18m14s & 1.494$\times10^{-3}$ & 7.6 & 7.61$^{+0.02}_{-0.01}$ & (5,4) \\ 
NGC~4395 & J1202.5+3332 & Seyfert 1.9 & 12h25m48.9s & +33d32m49s  & 1.064$\times10^{-3}$    & 3.9   & 5.6$^{+0.2}_{-0.1}$ & (3,5) \\ 
NGC~4941 & J1304.3-0532 & Seyfert 2   & 13h04m13.1s & -05d33m06s & 3.696$\times10^{-3}$     & 19.7  & 6.9$\pm0.3$ & (2,6)   \\
NGC~5273 & J1341.9+3537 & Seyfert 1.9 & 13h42m08.3s & +35d39m15s & 3.619$\times10^{-3}$     & 13.1  & 6.7$^{+0.1}_{-0.2}$ &  (2,7) \\ 
NGC~5643 & J1432.8-4412 & Seyfert 2   & 14h32m40.7s & -44d10m28s & 3.999$\times10^{-3}$     & 16.9  & 6.3$\pm0.4$ & (1,6) 
\enddata
\tablecomments{
Columns: (1) Galaxy name. (2) {\it Swift} ID. (3) Galaxy type taken from the {\it Swift}/BAT 70-month 
catalog \citep{Bau13}. 
(4)-(6) Position and redshift taken from the NASA/IPAC Extragalactic Database. 
(7) Luminosity distance in units of Mpc. (8) Black hole mass with 1$\sigma$ error. 
(9) References for distances and black hole masses. \\ 
{ References for distances.} \\ 
(1) \cite{Tul88}, (2) \cite{Theu07}, (3) \cite{Tul09}, (4) \cite{Spri09},  (5) \cite{Hum13} \\ 
{ References for black hole masses.} \\ 
(1) Mass calculated with the $M_{\rm BH}$-stellar velocity dispersion relation derived by 
\cite{Gul09}, where the stellar dispersion is taken from \cite{Ho09b}. 
(2) \cite{Mar15}, (3) \cite{Wal12},
(4) \cite{Herr99}, (5) \cite{Den15}, (6) \cite{Dav14},  (7)  \cite{Ben14}. 
}
\end{deluxetable*}

\section{OBSERVATION AND DATA REDUCTION}\label{sec:obs}

\subsection{Observations}

The basic information of our targets is summarized in Table~
\ref{tab:info_srcs}. The distances are taken from the literature or are 
calculated from the redshift. The black hole masses are compiled from the 
literature or are calculated from the stellar velocity dispersion by using 
the relation of \cite{Gul09}. {\it Suzaku} \citep{Mit07} observations of our 
targets were performed with exposures of $\sim$ 40--80 ksec. 
Table~\ref{tab:info_obs} gives the observation log.

{\it Suzaku} carries X-ray CCD cameras that cover the energy range below 
$\sim$ 10 keV called the X-ray Imaging Spectrometer (XIS), and a non-imaging 
instrument sensitive to hard X-rays above $\sim$ 15 keV called the hard X-ray 
Detector (HXD). Two front-illuminated XISs (FI-XISs; XIS-0 and XIS-3) and one 
back-illuminated XIS (BI-XIS; XIS-1) were available during the observations.
All targets were too faint above $\sim$ 50 keV to be detected with HXD/GSO. HXD/PIN was 
not operated during the observation of NGC 2655, and no HXD/PIN data of NGC 
4138 were available because of a telemetry saturation problem. NGC 2655, NGC 
3998, NGC 4258, NGC 4941, and NGC 5273 were observed at the XIS nominal 
position, whereas the others were observed at the HXD nominal position.

\begin{deluxetable}{ccccccc}
\tabletypesize{\small}
\tablecaption{Observation Log\label{tab:info_obs}}
\tablewidth{0pt}
\tablehead{ Target Name & Start Date of Obs. (UT) & Obs. ID \\ 
(1) & (2) & (3) 
}
\startdata
NGC~2655 & 2014 May 12 & 709003010 \\
NGC~3718 & 2009 Oct. 24 & 704048010 \\
NGC~3998 & 2014 May 03  & 709002010  \\
NGC~4102 & 2009 May 30 & 704057010 \\  
NGC~4138 & 2009 Nov. 02 & 704047010  \\
NGC~4258 & 2010 Nov. 11 & 705051010  \\
NGC~4395 & 2007 June 02 &  702001010 \\
NGC~4941 & 2012 June 22 & 707001010  \\
NGC~5273 & 2013 July 16 & 708001010  \\
NGC~5643 & 2007 Aug. 19 & 702010010 
\enddata
\tablecomments{
Columns: (1) Galaxy name.
(2) Observation start date.
(3) {\it Suzaku} observation ID.
}
\end{deluxetable}

FTOOLS (version 6.15.1) and XSPEC (version 12.8.1.g) are used for the data 
reduction and spectral analysis. The CALDB released on 2015 Jan 5 are 
utilized. We reprocess the XIS event files according to the ABC 
guide
and extract source events from a circular region 
centered on the target centroid. The background events are taken from an 
off-source region within the XIS field-of-view. The FI-XISs data are merged 
to increase the signal-to-noise ratio. We utilize the cleaned event files of 
HXD/PIN provided by the {\it Suzaku} team. To reproduce the background 
spectra, we utilize the ``tuned'' Non X-ray background model \citep{Fuka09} 
and the Cosmic X-ray background spectrum simulated by assuming the model given
 in \cite{Gru99}. From eight targets except NGC 2655 and NGC 4138, we detect
significant signals with HXD/PIN well above the systematic uncertainties in 
the Non X-ray background model (1\% for the exposure time above 40 ksec; 
\citealt{Fuka09}). In the spectral analysis, we also use the {\it Swift}/BAT 
spectra averaged for 70 months \citep{Bau13}. 


\subsection{Light Curves}\label{sec:lc}

\begin{figure*}[!ht]
\begin{center} \vspace{-1cm}
\includegraphics[scale=0.3,angle=-90]{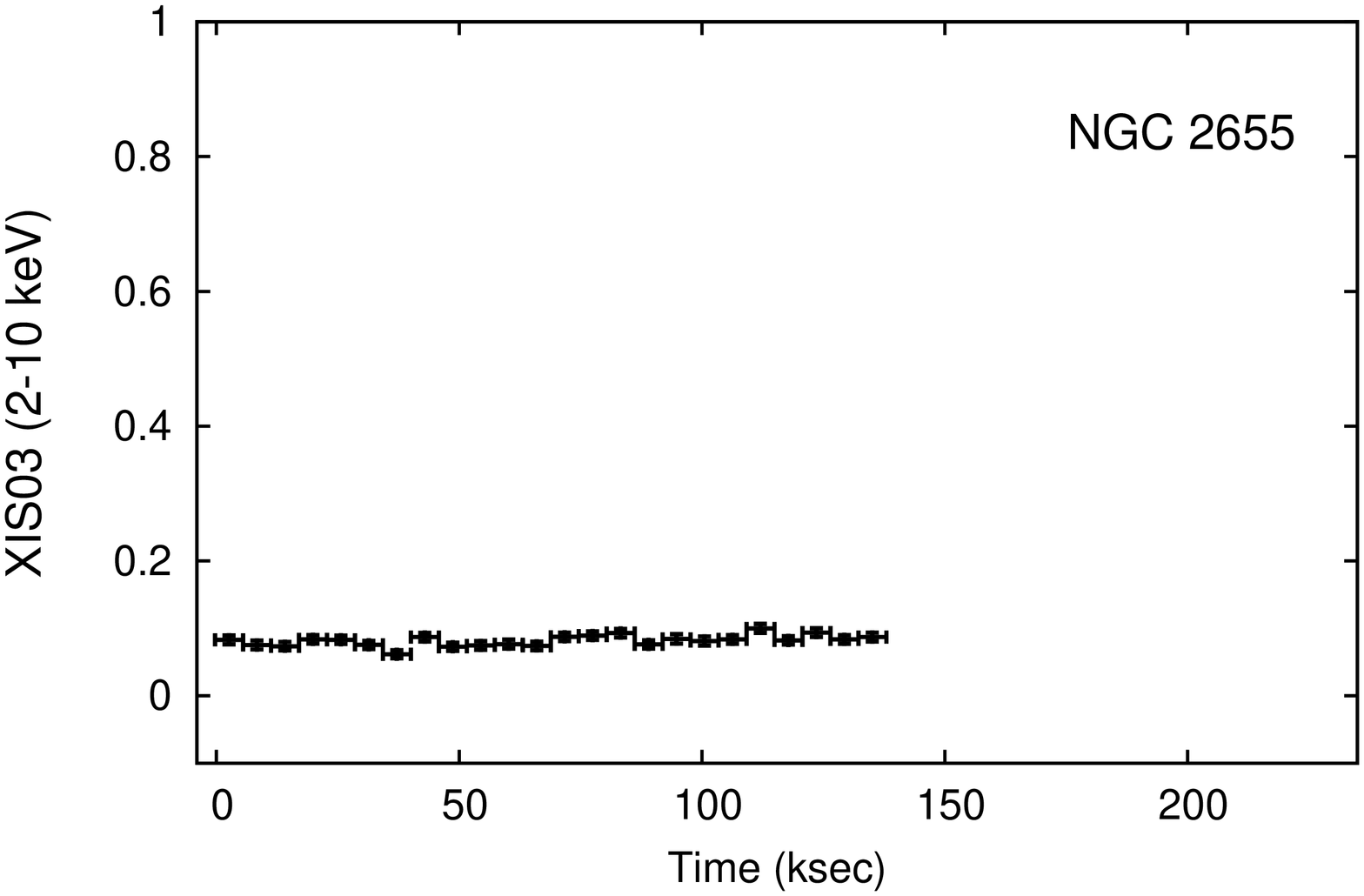}  \hspace{-0.5cm}
\includegraphics[scale=0.3,angle=-90]{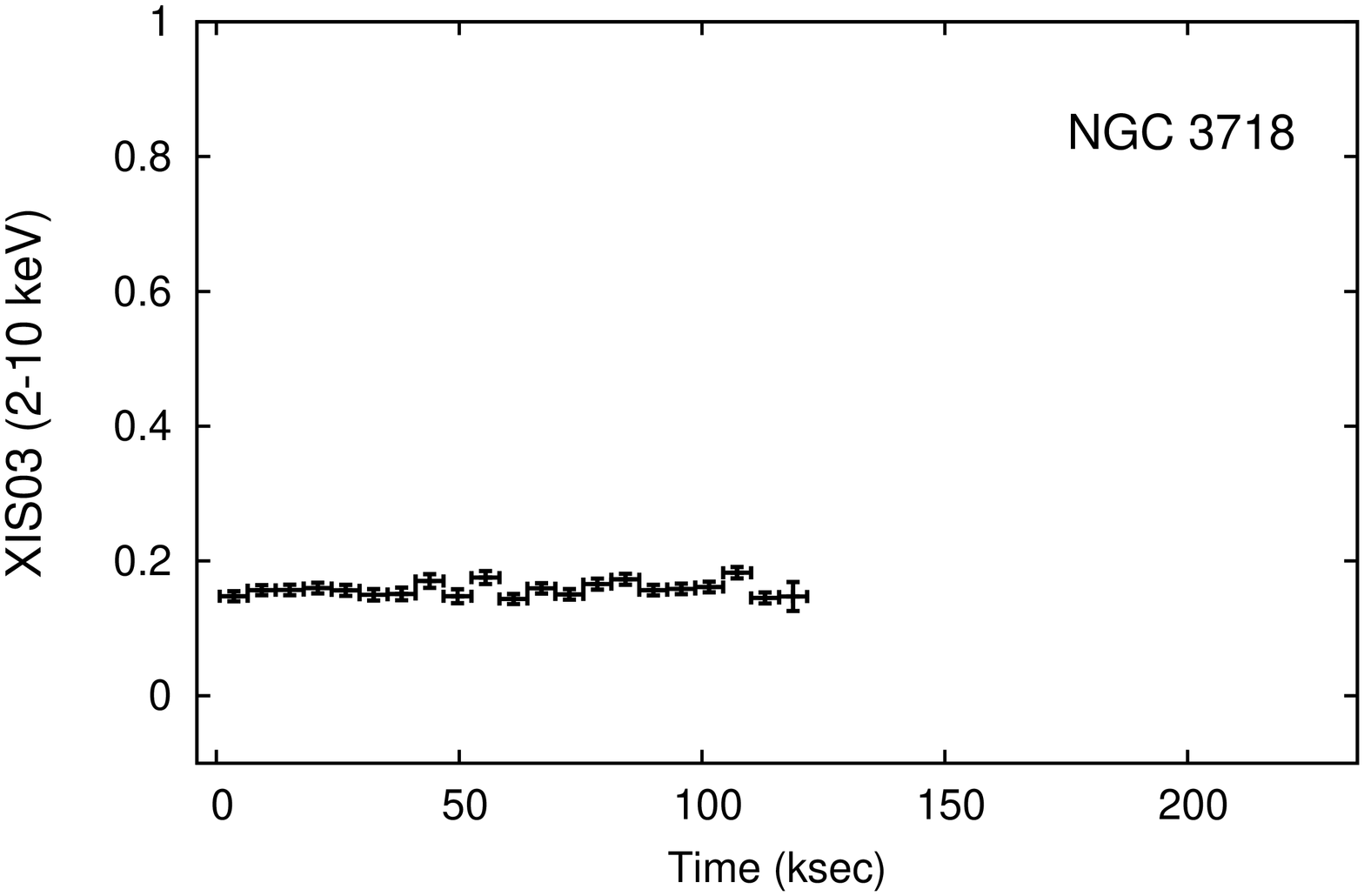}  \\ \vspace{-0.5cm} 
\includegraphics[scale=0.3,angle=-90]{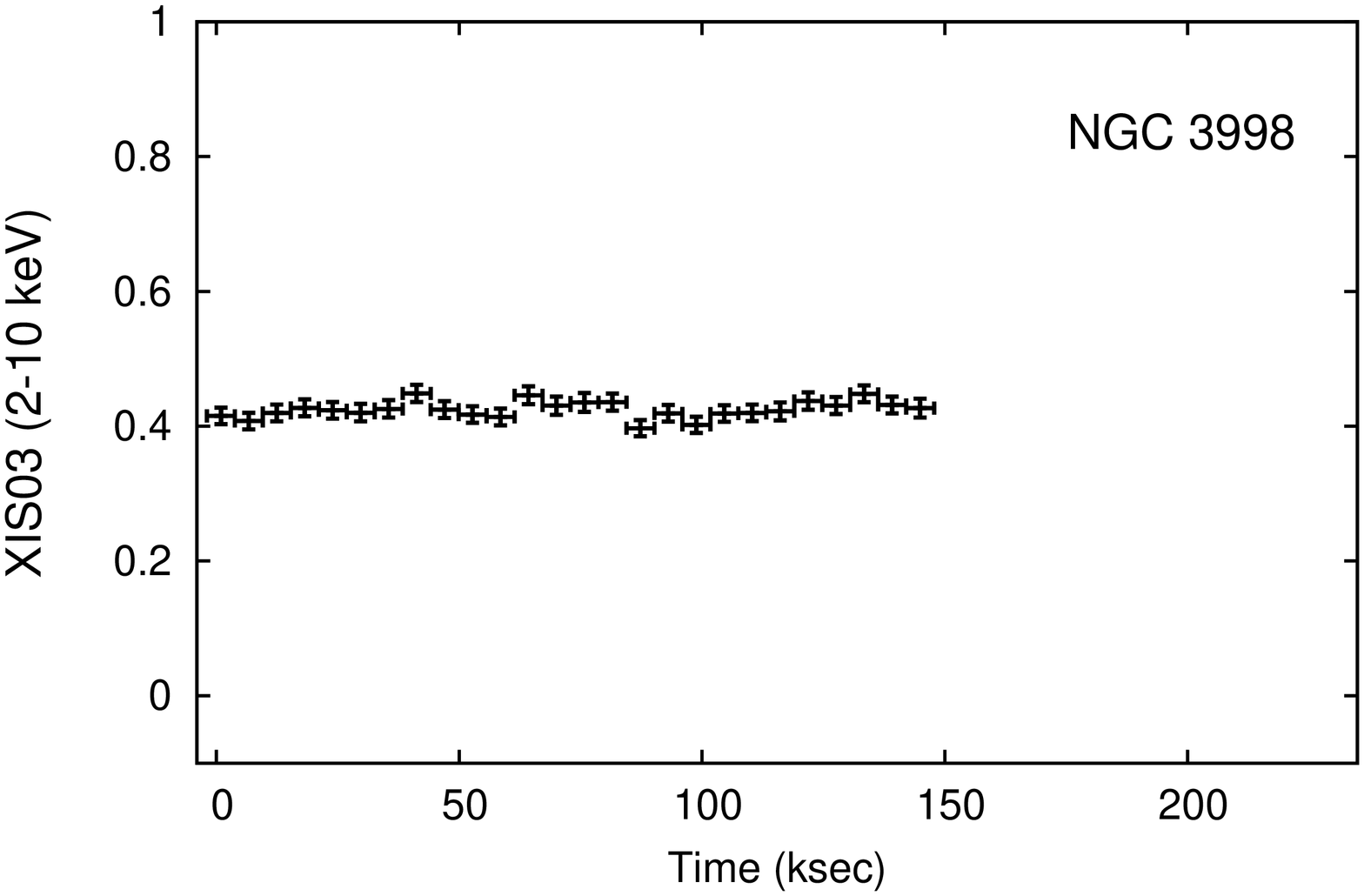}  \hspace{-0.5cm}
\includegraphics[scale=0.3,angle=-90]{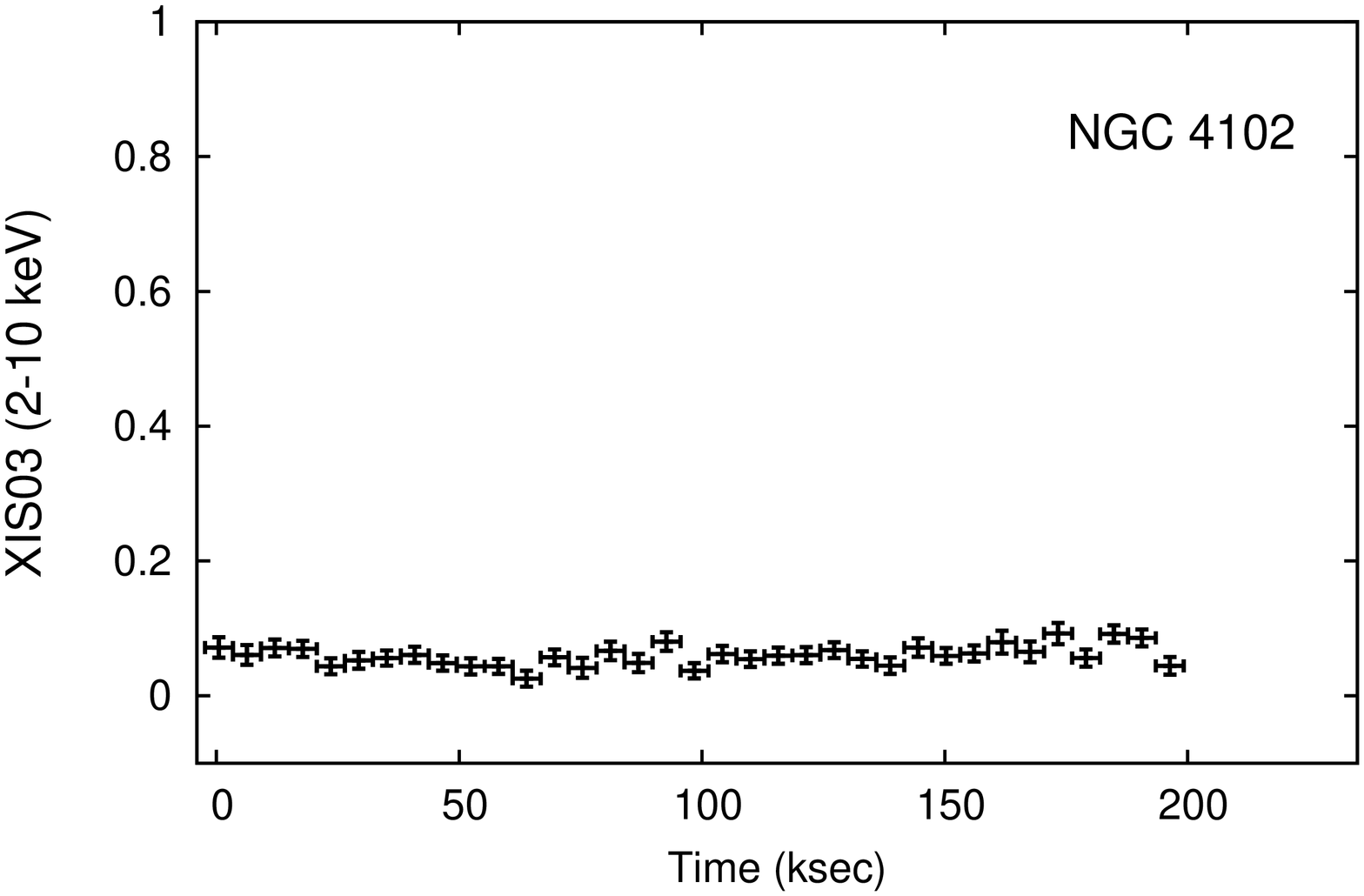} \\  \vspace{-0.5cm} 
\includegraphics[scale=0.3,angle=-90]{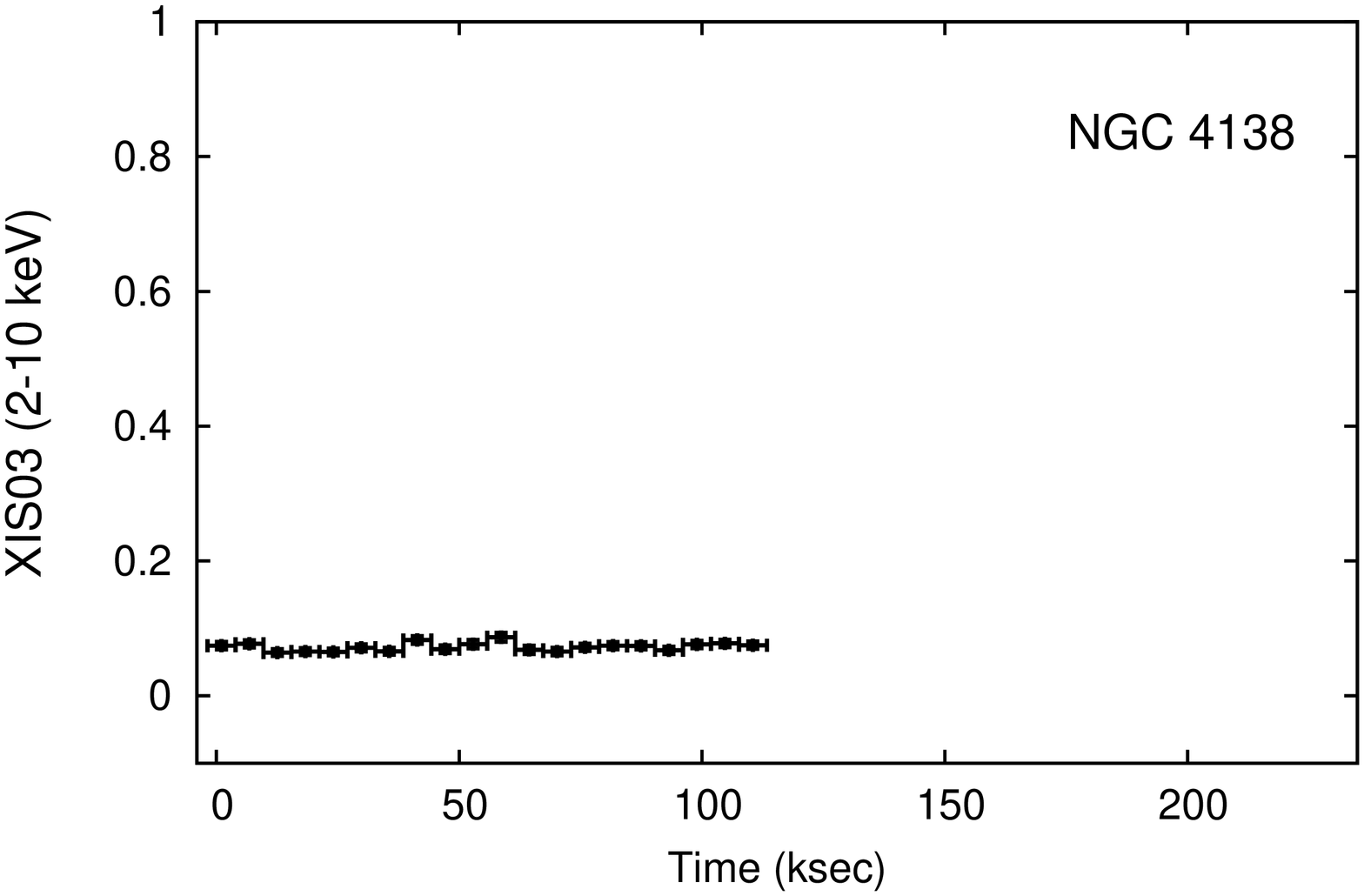} \hspace{-0.5cm}
\includegraphics[scale=0.3,angle=-90]{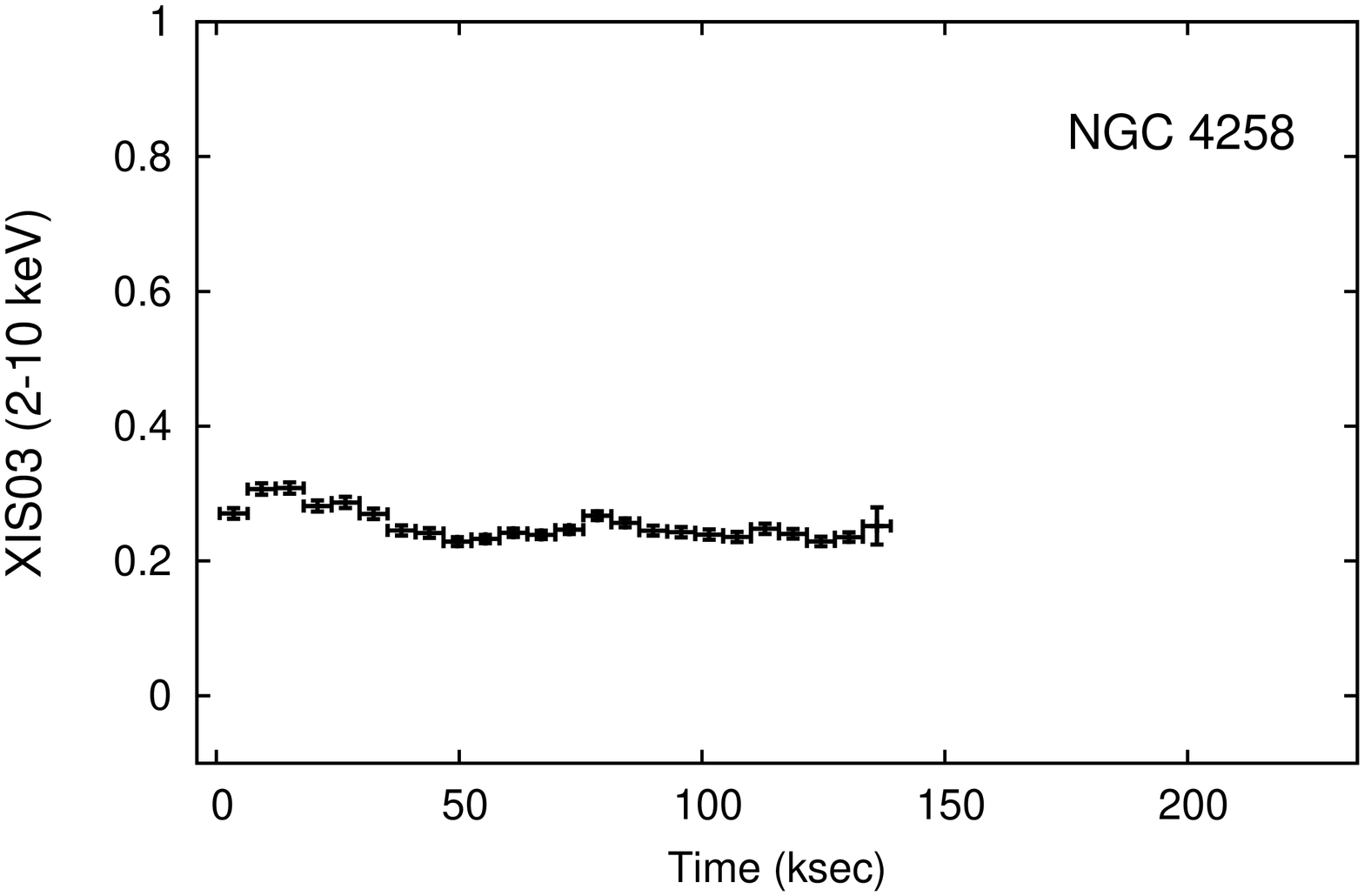} \\  \vspace{-0.5cm} 
\includegraphics[scale=0.3,angle=-90]{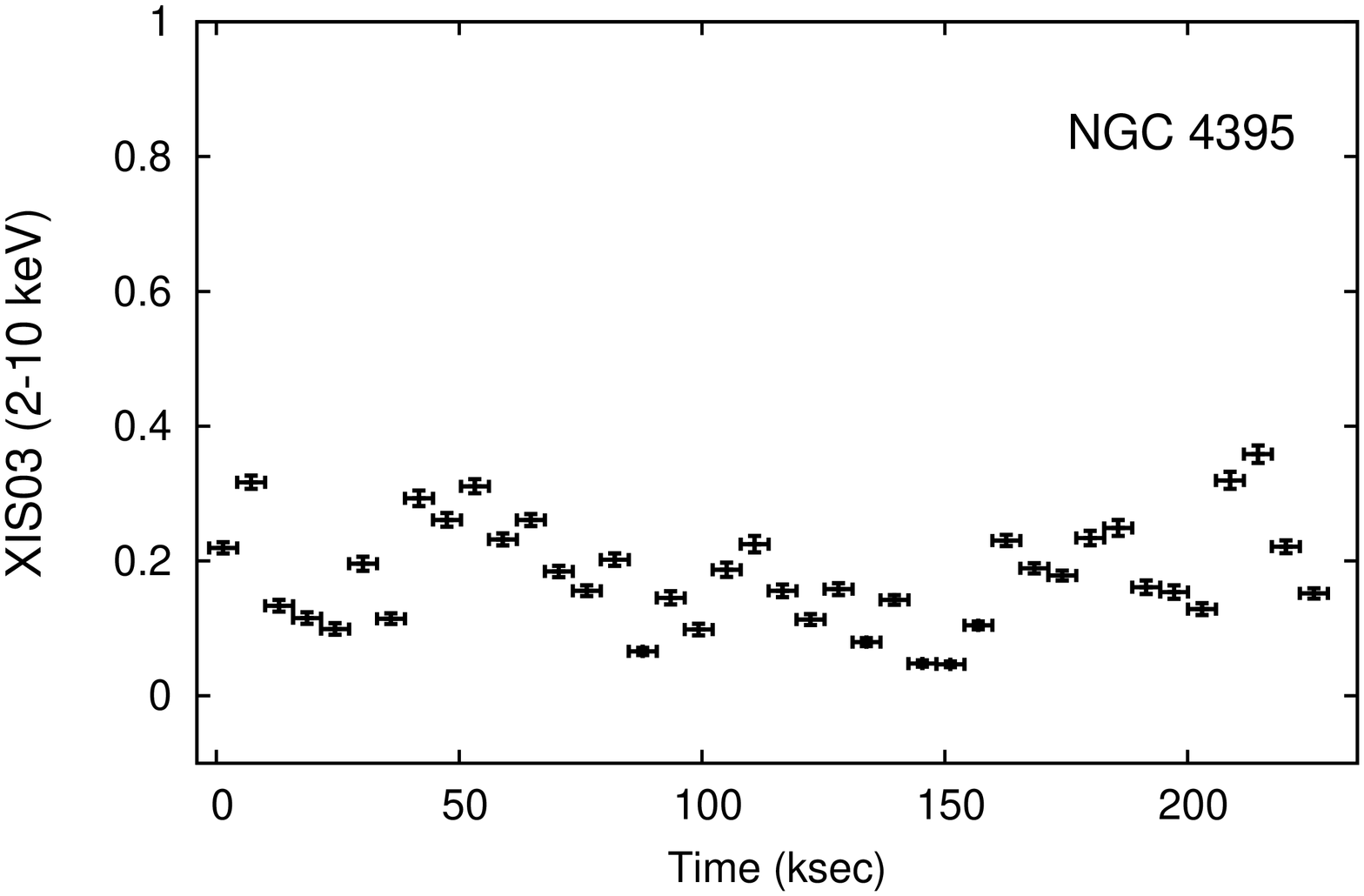} \hspace{-0.5cm}
\includegraphics[scale=0.3,angle=-90]{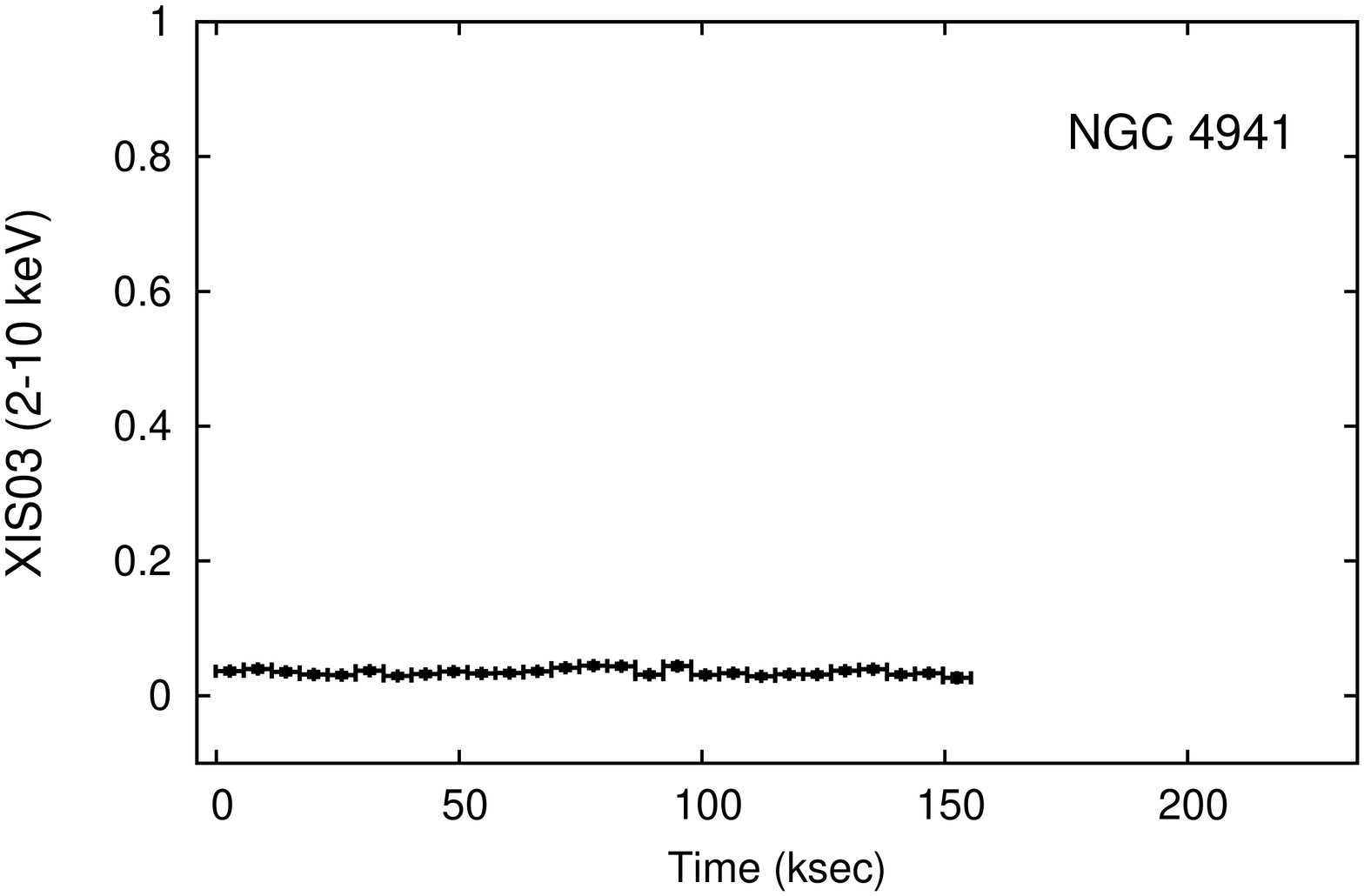} \\  \vspace{-0.5cm} 
\includegraphics[scale=0.3,angle=-90]{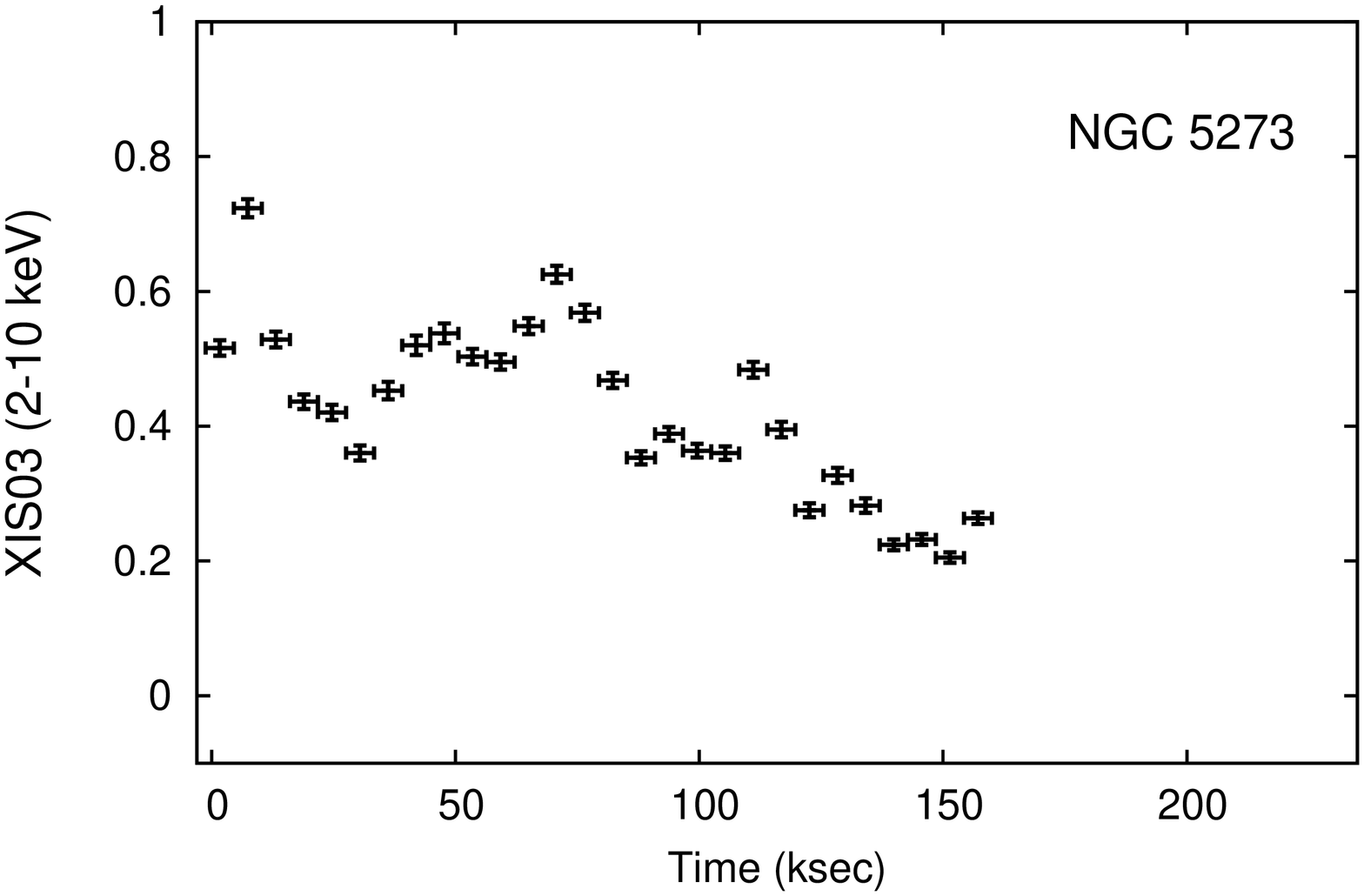} \hspace{-0.5cm}
\includegraphics[scale=0.3,angle=-90]{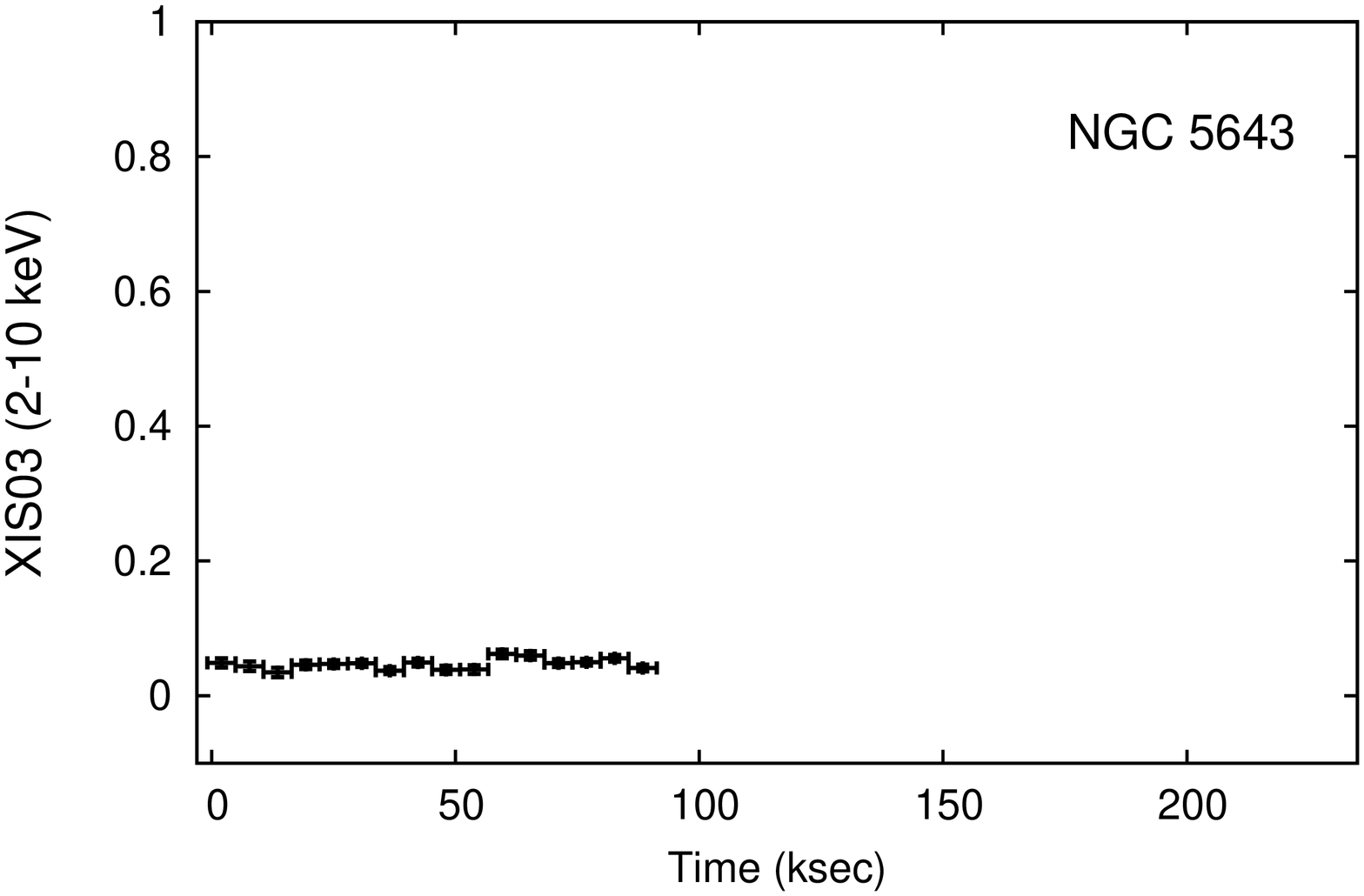} 
\caption{
Count-rate light curves in the 2--10 keV band obtained with FI-XISs. 
}
\label{fig:lc}
\end{center}
\end{figure*}

Figure~\ref{fig:lc} shows the background-subtracted light curves of our sample
 in the 2--10 keV band (XIS-0+XIS-3). The bin size is set to 5760 sec 
(the orbital period of {\it Suzaku}) to exclude any modulations that depend on 
the orbital phase. Strong time variability is seen in NGC~4395 and NGC~5273. 
To quantify it, we calculate 
the mean fractional variation ($F_{\rm var}$) defined in equation (3) of 
\cite{Rod97}: 
\begin{eqnarray}
 F_{\rm var} = \sqrt{\frac{S^2-\overline{\sigma^2}}{\overline{r}^2}},     \nonumber 
\end{eqnarray}
where 
\begin{eqnarray}
  S^2 = \frac{1}{N-1} \sum^{N}_{i=1} (r_i - \overline{r})^2 \nonumber 
\end{eqnarray} 
and 
\begin{eqnarray}
 \overline{\sigma^2} = \frac{1}{N} \sum^{N}_{i=1} \sigma^2_i.     \nonumber 
\end{eqnarray} 
Here, for the sample size $N$, $r$ and $\overline{r}$ is an observed flux 
and the mean value, respectively, while $\sigma$ an error.
Larger $F_{\rm var}$ means that the object is more variable. Also, 
by fitting a constant flux model to the light curves, we calculate the 
probability of obtaining the observed data  (p-value) based on chi-square 
distribution. $F_{\rm var}$ and p-values of the 2--10 keV light curves are 
summarized in Table~\ref{tab:info_val}. 
Judging from $F_{\rm var} > 0.3$ and p-value $< 0.01$, NGC~4395 and 
NGC~5273 were highly variable. 
This would reflect the small spatial scale of the X-ray emitting region 
inferred from their low SMBH masses ($M_{\rm BH} < 5\times10^{6} M_{\rm sol}$, 
where $M_{\rm sol}$ is the solar mass).  Although the SMBH mass of NGC 5643 is 
also low, its variability may be suppressed due to heavy obscuration of the 
direct transmitted emission (\citealt{Gua04}; \citealt{Mat13}). 
As we will discuss in 
Section~\ref{sec:broad_ana_2}, in these low-mass LLAGNs
the flux of the reflection component from the torus follows
that of the primary emission on time scales shorter than 70 months.

\begin{deluxetable}{p{3cm}p{2cm}cccc}
\tabletypesize{\small}
\tablecaption{Variability in the 2--10 keV Light Curves\label{tab:info_val}}
\tablewidth{0pt}
\tablehead{ Target Name & $F_{\rm var}$ &  p-value \\ 
(1) & (2) & (3) 
}
\startdata
NGC~2655 & 0.064 & 0.02 \\ 
NGC~3718 & 0.031 & 0.08 \\ 
NGC~3998 & 0.003 & 0.40 \\ 
NGC~4102 & 0.138 & 0.07 \\
NGC~4138 & 0.040 & 0.24 \\ 
NGC~4258 & 0.083 & $< 0.01$ \\ 
NGC~4395 & 0.424 & $< 0.01$ \\ 
NGC~4941 & 0.090 & 0.02 \\ 
NGC~5273 & 0.303 & $< 0.01$ \\ 
NGC~5643 & 0.106 & 0.06 
\enddata
\tablecomments{
Columns: (1) Galaxy name.
(2) Fractional variability. 
(3) Probability of obtaining the observed light curve at constant flux.
}
\end{deluxetable}

\section{ANALYSIS AND RESULTS}\label{sec:spec_ana}

\subsection{Broadband Spectral Analysis}\label{sec:broad_ana}

\begin{figure*}[!ht]
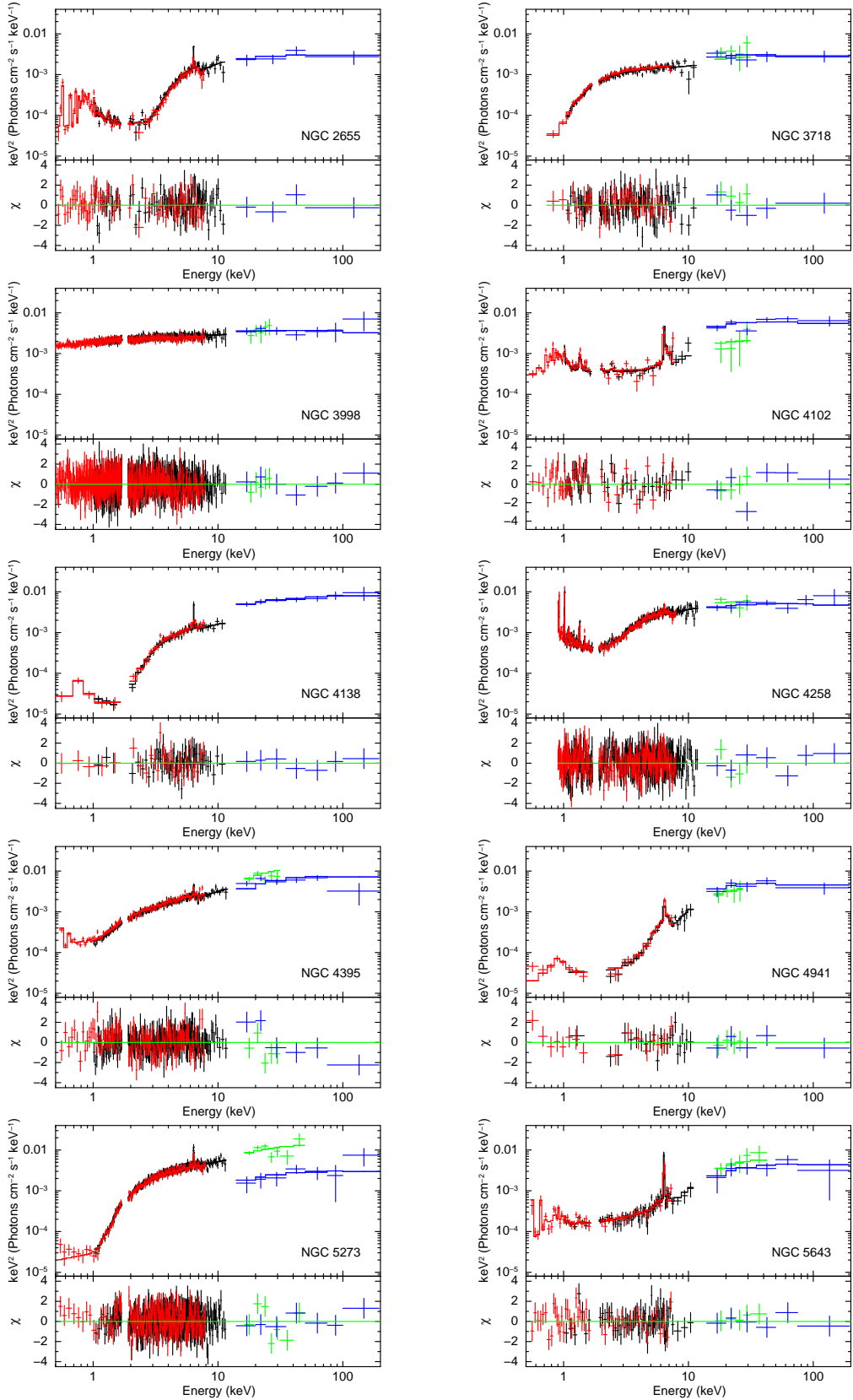

\centering
\includegraphics[scale=0.24,angle=-90]{outfigures_ngc_2655.ps} \hspace{1cm}
\includegraphics[scale=0.24,angle=-90]{outfigures_ngc_3718.ps} \\ \vspace{-0.3cm}
\includegraphics[scale=0.24,angle=-90]{outfigures_ngc_3998.ps} \hspace{1cm}
\includegraphics[scale=0.24,angle=-90]{outfigures_ngc_4102.ps} \\ \vspace{-0.3cm}
\includegraphics[scale=0.24,angle=-90]{outfigures_ngc_4138.ps} \hspace{1cm}
\includegraphics[scale=0.24,angle=-90]{outfigures_ngc_4258.ps} \\ \vspace{-0.3cm}
\includegraphics[scale=0.24,angle=-90]{outfigures_ngc_4395.ps} \hspace{1cm}
\includegraphics[scale=0.24,angle=-90]{outfigures_ngc_4941.ps} \\ \vspace{-0.3cm}
\includegraphics[scale=0.24,angle=-90]{outfigures_ngc_5273.ps} \hspace{1cm}
\includegraphics[scale=0.24,angle=-90]{outfigures_ngc_5643.ps}  \\
\caption{
Unfolded {\it Suzaku} and {\it Swift}/BAT spectra. 
In the upper panels, the unfolded spectra of 
FI-XISs, BI-XIS, HXD/PIN, and {\it Swift}/BAT are represented with the black, red, green, and blue crosses, 
respectively. The solid curves show the best-fit model. In the lower panels, the 
residuals are shown. 
}
\label{fig:unf_spec}
\end{figure*}

We simultaneously analyze the X-ray spectra of the FI-XISs, BI-XIS, HXD/PIN, 
and {\it Swift}/BAT, which cover the 1--12 keV, 0.5--8 keV, 14--60 keV, and 
15--200 keV bands, respectively. We exclude the 1.7--1.9 keV band of the XIS 
spectra to avoid systematic uncertainties of the energy response around the 
Si K-edge region.  The cross-normalization factor of the HXD/PIN spectrum to 
the FI-XISs one is set to 1.16 (1.18) for the XIS (HXD) nominal position 
observation, and that of the BI-XIS spectrum is allowed to vary. 

On the basis of the previous works (\citealt{Taz13}; \citealt{Kawa13}), we 
start with a base-line model,  
\begin{small}
\begin{eqnarray}
& &  \tt constant*zphabs*zpowerlw*zhighect \nonumber \\
& & \tt  +constant*zpowerlw*zhighect+pexrav+zgauss \nonumber 
\end{eqnarray}
\end{small} 
in the XSPEC terminology. This model includes absorbed primary X-ray emission 
(i.e., a cut-off power law), a scattered component, and a reflection 
continuum from distant cold matter accompanied with a narrow iron-K$\alpha$ 
line. Optically-thin thermal emission from the host galaxy ({\tt apec} in 
XSPEC) and other emission/absorption lines ({\tt zgauss}) are also added if 
they are significantly required with a confidence level above 90\% in terms 
of $\Delta \chi^2$. Because it is difficult to determine the cut-off energy 
from our data, we fix it at 300 keV, a typical value measured in nearby AGNs 
\citep{Dad08}. The first {\tt constant} factor, $N_{\rm XIS}$, is applied to 
the primary power-law component in the {\it Suzaku} spectra to absorb 
possible time variability between the {\it Suzaku} (one epoch) and 
{\it Swift}/BAT (averaged for 70 months) observations. The second 
{\tt constant} term represents the scattered fraction, $f_{\rm scatt}$. As a 
reflection component from the torus, we employ the {\tt pexrav} model, which 
calculates a reflected spectrum from an optically thick slab with a solid 
angle of $\Omega$ irradiated by a point source \citep{Mag95}. We set the 
reflection strength, $R = \Omega/2\pi$, as a free parameter, and fix the 
inclination angle at 60$^\circ$. 
It is  confirmed that even if 30$^\circ$ is adopted, best-fit parameters do not 
significantly change. 
The shape of the incident spectrum is assumed 
 to be the same as the power-law component. We basically assume that the 
reflection and scattered components did not vary between the {\it Suzaku} and 
{\it Swift}/BAT observations, considering that the size of the reflector has 
most likely a pc scale. In the three low mass LLAGNs (NGC 4395, NGC 5273, and 
NGC 5643), however, we assume that the reflection component varied in 
accordance with the primary emission because of a smaller size of the 
emitting regions. Thus, in these targets, $R$ is defined 
with respect to the primary component in the {\it Suzaku} data. 
The {\tt zgauss} component represents an
iron-K$\alpha$ fluorescence line. The line width is 
fixed at $20$ eV, which corresponds to a typical velocity dispersion 
of $\sim$ 2000 km s$^{-1}$  measured 
with {\it Chandra}/HETGS in local Seyfert galaxies \citep{Shu10}. 
We always consider the Galactic absorption $(N^{\rm gal}_{\rm H})$, 
which is calculated with the {\tt nh} command \citep{Kal05} in FTOOLS.

\begin{figure*}[!ht]
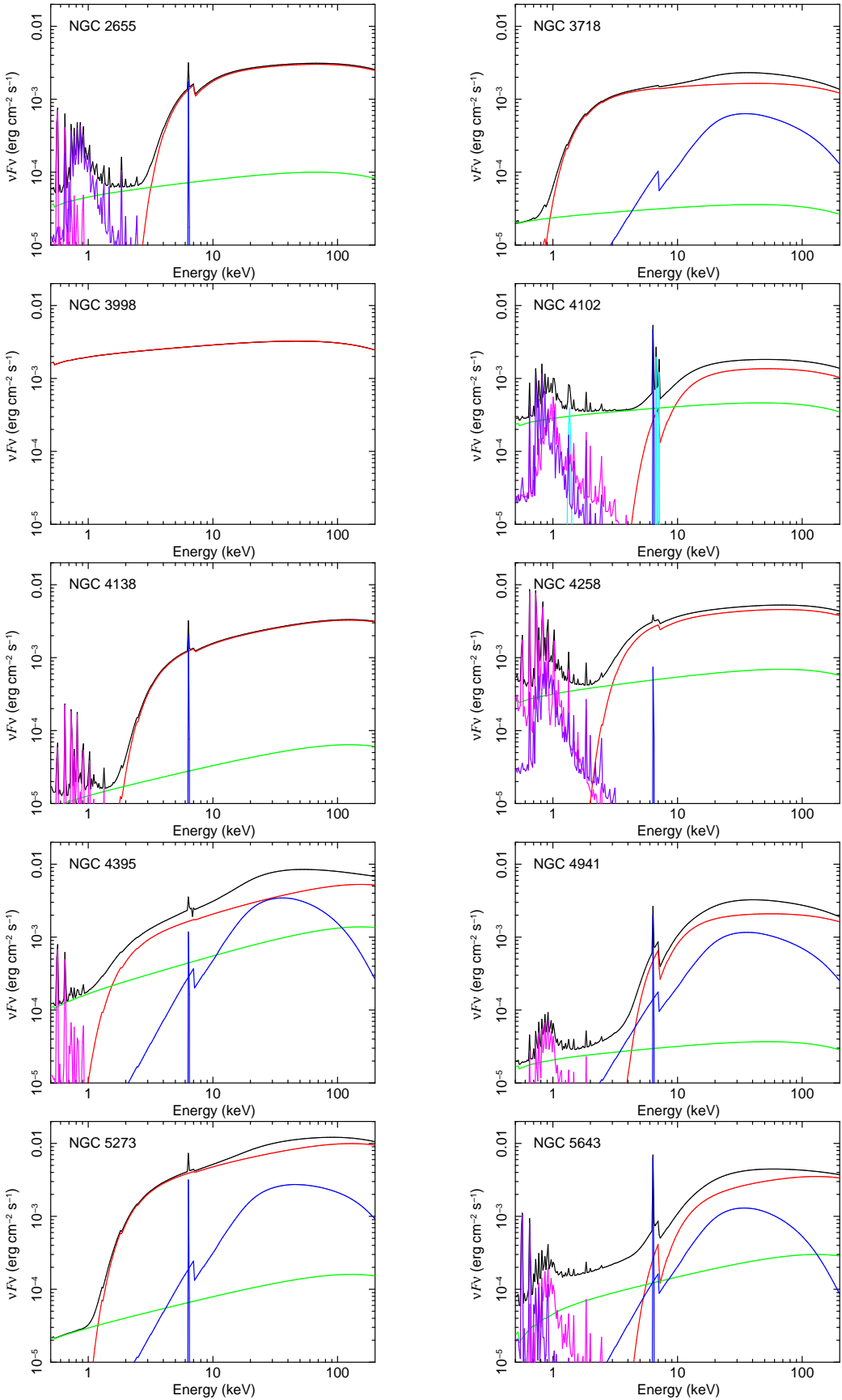

\centering
\includegraphics[scale=0.25,angle=-90]{outfigures_model_ngc_2655.ps} \hspace{1cm}
\includegraphics[scale=0.25,angle=-90]{outfigures_model_ngc_3718.ps} \\ 
\includegraphics[scale=0.25,angle=-90]{outfigures_model_ngc_3998.ps} \hspace{1cm}
\includegraphics[scale=0.25,angle=-90]{outfigures_model_ngc_4102.ps} \\
\includegraphics[scale=0.25,angle=-90]{outfigures_model_ngc_4138.ps} \hspace{1cm}
\includegraphics[scale=0.25,angle=-90]{outfigures_model_ngc_4258.ps} \\
\includegraphics[scale=0.25,angle=-90]{outfigures_model_ngc_4395.ps} \hspace{1cm}
\includegraphics[scale=0.25,angle=-90]{outfigures_model_ngc_4941.ps} \\ 
\includegraphics[scale=0.25,angle=-90]{outfigures_model_ngc_5273.ps} \hspace{1cm}
\includegraphics[scale=0.25,angle=-90]{outfigures_model_ngc_5643.ps} 
\caption{
Best-fit models in units of $\nu F\nu$. The black, red, green, blue, and cyan 
lines correspond to the total, transmitted emission, scattered component, 
reflection components from the torus, and other emission lines than the 
iron-K$\alpha$ line, respectively. The purple and magenta lines represent the 
optically-thin thermal emission. 
}
\label{fig:unf_model}
\end{figure*}

The details of the spectral analysis for individual targets are described in 
Section~\ref{sec:broad_ana_2}. The best-fit parameters, observed fluxes, and 
absorption corrected luminosities are summarized in Tables~\ref{tab:para} and 
\ref{tab:flux}. The unfolded spectra and best-fit models are plotted in 
Figure~\ref{fig:unf_spec} and \ref{fig:unf_model}. 
The Eddington ratios in Table~\ref{tab:flux} 
are calculated from the luminosity in the 2--10 keV band measured with {\it Suzaku} 
({\it Swift}) for the low (high) mass LLAGNs, by assuming a bolometric correction factor 
of 10 (\citealt{Ho09}; \citealt{Vas09}).

\subsection{Notes on Individual Objects}
\label{sec:broad_ana_2}

\subsubsection{NGC 2655}

For NGC 2655, the base-line model alone does not give an acceptable fit, 
leaving residuals in the soft X-ray band below 2 keV. Although one {\tt apec} 
component is insufficient to improve the fit ($\chi^2/d.o.f = 226.4/169$), 
adding a second {\tt apec} model gives an acceptable fit 
($\chi^2/d.o.f = 186.6/167$). We obtain a hydrogen column density of 
$N_{\rm H} = 2.61^{+0.27}_{-0.17} \times 10^{23}$ cm$^{-2}$ and 
$\Gamma = 1.77^{+0.19}_{-0.07}$, which are consistent with the {\it ASCA} 
result by \cite{Tera02}. The temperatures of the {\tt apec} models are 
$0.18^{+0.06}_{-0.05}$ keV and $0.74 \pm 0.04$ keV. The higher temperature is 
consistent with the {\it XMM-Newton} result \citep{Gon09}. Although the lower 
temperature component has not been reported for this source, the value is 
within a range observed in nearby Seyfert galaxies (e.g., \citealt{Cap06}).

We significantly detect an iron-K$\alpha$ line at 6.4 keV, which was not 
detected in the previous {\it ASCA} and {\it XMM-Newton} observations 
(\citealt{Tera02}; \citealt{Gon09}). The iron-K$\alpha$ line fluxes observed 
with {\it ASCA}, {\it XMM-Newton}, and {\it Suzaku} were 
$< 4.9\times10^{-6}$ photons s$^{-1}$ cm$^{-2}$, $< 3.3\times10^{-6}$ photons 
s$^{-1}$ cm$^{-2}$, and $5.3\pm 1.2\times10^{-6}$ photons s$^{-1}$ cm$^{-2}$, 
respectively. Hence, the iron line flux slightly increased in the {\it Suzaku}
observation (performed in 2014) compared with the {\it XMM-Newton} epoch 
(2005).  This indicates that the spatial size of the reflector (torus) is 
less than a few pc.

\subsubsection{NGC 3718}

The base-line model well reproduces the X-ray spectra of NGC 3718, yielding 
$\chi^2/d.o.f = 214.9/206$ with $N_{\rm H} = 1.29^{+0.12}_{-0.11} \times 10^{22}$ 
cm$^{-2}$ and $\Gamma = 1.86^{+0.13}_{-0.09}$. No significant iron-K$\alpha$ 
line is detected. \cite{Her14} also reported a similar result by analyzing 
the {\it Chandra} and {\it XMM-Newton} spectra.

\subsubsection{NGC 3998} 

The base-line model without a scattering component well reproduces the spectra with 
$\chi^2/d.o.f$ = 950.8/982. The obtained photon index, 
$\Gamma = 1.84\pm0.01$, is consistent with a previous study 
\citep{Ptak04}. We find that the reflection strength is very weak with an 
upper limit of $R_{\rm torus} < 0.10$. No significant iron-K$\alpha$ line 
emission is detected, either. These suggest that there is little surrounding 
matter around the nucleus.

\subsubsection{NGC 4102}

The spectra of NGC 4102 are complex, and cannot be fit with the base-line model
 alone. By including two apec models, we obtain a better fit with 
$\chi^2/d.o.f = 348.15/271$. Further, adding three narrow emission lines other
 than the neutral iron-K$\alpha$ line largely improves the fit 
($\chi^2/d.o.f = 299.3/265$). Each line component improves the fit by 
$\Delta \chi^2 > 7$ for two degrees of freedom (center energy and 
normalization). We obtain the line energies of $1.36\pm0.02$ keV, 
$6.75^{+0.03}_{-0.04}$ keV, and $7.11^{+0.18}_{-0.10}$ keV, which possibly 
correspond to Mg XI, slightly blue-shifted Fe XXV, and Fe K$\beta$ lines, 
respectively. The reflection strength derived from the continuum fit is small,
 $R_{\rm torus} < 0.14$, whereas the observed equivalent width of the 
iron-K$\alpha$ line is large ($\sim 900$ eV). This implies that the reflector
 may not be Compton thick. 

\subsubsection{NGC 4138}

The base-line model combined with a single {\tt apec} model well reproduces 
the X-ray spectra of NGC 4138 with $\chi^2/d.o.f = 78.1/94$. The addition of 
the {\tt apec} model improves the fit by $\Delta \chi^2 = 26.9$ for two 
degrees of freedom (temperature and normalization).  We obtain 
$N_{\rm H} =7.82^{+0.72}_{-0.68} \times10^{22}$ cm$^{-2}$ and 
$\Gamma = 1.59^{+0.17}_{-0.06}$, which are consistent with the result of the 
{\it XMM-Newton} observation in 2001 \citep{Cap06}. 

\subsubsection{NGC 4258}

The BI-XIS spectrum of NGC 4258 below 0.9 keV shows highly complex line 
features, and therefore we exclude this energy band for our analysis. The 
base-line model combined with two {\tt apec} models gives a good fit 
($\chi^2/d.o.f = 640.8/573$), which is better than that with a single
{\tt apec} model ($\chi^2/d.o.f = 659.1/575$). The obtained $N_{\rm H}$,
$\Gamma$, and temperatures of the {\tt apec} components are consistent with 
the results by \cite{Mak94} based on the {\it ASCA} observation in 1993. A 
significant narrow iron-K$\alpha$ line at 6.4 keV is detected in our {\it Suzaku} 
spectra. The observations with {\it Chandra} in 2000 and 2001 and 
{\it XMM-Newton} in 2006 revealed that the observed iron-K$\alpha$ line fluxes
 were variable (e.g., \citealt{You04}; \citealt{Rey09}). Indeed, the 
{\it Chandra} spectra did not show the iron-K$\alpha$ emission line 
\citep{You04}.  Also, analyzing the {\it Suzaku} and {\it XMM-Newton} spectra,
 \cite{Rey09} concluded that the narrow iron-K$\alpha$ emission was 
originated from the accretion disk because of its flux variability on time 
scales of 160 days. 
Hence, the iron line emitting region of NGC 4258 may not 
be distant cold matter or a dusty torus. 
This fact is also recognized when we statistically discuss the 
torus structure of LLAGNs in Section~\ref{sec:lumEdd2iron}.

\subsubsection{NGC 4395}

The spectra of NGC 4395 are well reproduced with the base-line model plus a 
narrow absorption line at $6.90^{+0.08}_{-0.07}$ keV 
($\chi^2/d.o.f = 332.7/354$). The absorption line improves the fit by 
$\Delta \chi^2 = 7.5$, and is consistent with Fe XXVI K$\alpha$. We obtain 
$R_{\rm torus} = 1.67^{+1.73}_{-0.92}$ by assuming that the fluxes of the 
reflected emission varied between the {\it Suzaku} and {\it Swift}/BAT spectra
 (see above). We note that without this assumption we would derive an 
unphysically large reflection strength, $R_{\rm torus} \sim 3.5$, which is 
much larger than typical values of $R_{\rm torus}\sim1$ observed in local AGNs 
\citep{Dad08}.

\subsubsection{NGC 4941}

The base-line model plus an {\tt apec} model well reproduces the spectra with 
$\chi^2/d.o.f = 33.1/43$.  The broadband X-ray spectra were already analyzed 
by \cite{Kawa13}, and the difference from the previously used model is the 
absence of the disk-reflection component. The obtained parameters are 
consistent with each other.

\subsubsection{NGC 5273}

The base-line model gives a good fit with $\chi^2/d.o.f$ = 586.4/563. The 
observed equivalent width of the iron-K$\alpha$ line, $\sim 90$ eV, is smaller than 
that estimated from the {\it XMM-Newton} data taken in 2002, $\sim$ 230 eV 
\citep{Cap06}, even though the observed continuum flux are almost the same. 
This suggests that the line emitting region must be smaller than a few pc and 
that the averaged past activity before each observation, which determines the
iron-K$\alpha$ line flux, was higher in the earlier epoch. Similarly to the case of 
NGC 4395, we would obtain an unphysically large reflection strength of 
$R_{\rm torus} \sim 5.7$ if we assumed that the reflection component did not 
change between the {\it Suzaku} and {\it Swift}/BAT observations.

\subsubsection{NGC 5643}

{\it XMM-Newton} observations (\citealt{Gua04}; \citealt{Mat13}) showed that 
NGC 5643 contains a ULX, which is located at $\approx$ 0.9' from the nucleus 
and cannot be resolved with the {\it Suzaku} beamsize ($\approx 2'$). Because 
the nucleus is heavily obscured, the ULX emission may contaminate the XIS 
spectra at energies below 10 keV. Hence, we take into account the ULX spectrum
 (see Appendix) in our spectral analysis. 

The base-line model plus the ULX component and two {\tt apec} models well 
reproduces the spectra ($\chi^2/d.o.f$ = 109.5/114).  The large hydrogen 
column density of $N_{\rm H} = 9.4^{+6.1}_{-3.2}\times 10^{23} $ cm$^{-2}$ and
 strong reflection strength of $R_{\rm torus} \sim 0.9$ confirm that NGC 5643 
has a heavily obscured AGN. We obtain an (apparent) scattered fraction of 
$f_{\rm scatt} = 10.1^{+23.2}_{-7.4}\%$, although it strongly depends on the 
assumed flux level of the ULX. When we increase the ULX flux by a factor of 2 
to make it consistent with that measured by the {\it XMM-Newton} observation 
in 2003 \citep{Gua04}, we obtain only an upper limit of 
$f_{\rm scatt} \lesssim 1\%$. 

Our data can be used to constrain the variability of the ULX if we assume 
that the AGN is less variable. The summed 2--10 keV flux from the nucleus and 
ULX as observed with {\it Suzaku} in 2007 was $1.3\times10^{-12}$ erg s$^{-1}$
 cm$^{-2}$, while the fluxes from the nucleus measured with {\it XMM-Newton} 
in 2003 and 2009 were both $\approx 8\times10^{-13}$ erg s$^{-1}$ cm$^{-2}$ 
(\citealt{Gua04}; \citealt{Mat13}). By assuming that the flux from the nucleus
 remained constant, the ULX flux during the {\it Suzaku} observation is 
estimated to be $\approx 5\times10^{-13}$ erg s$^{-1}$. This value is between 
that during the {\it XMM-Newton} observation in 2003 ($7.5\times10^{-13}$ erg
s$^{-1}$ cm$^{-2}$) and that in 2009 ($\sim 3.5\times10^{-13}$ erg s$^{-1}$ 
cm$^{-2}$), implying a consistent decay of the ULX flux from 2003 to 2009.


\tabletypesize{\scriptsize}
\begin{deluxetable*}{cccccccccccccccccccccccc}
\tablecaption{Best-fit Parameters obtained from broadband X-ray spectral analysis
\label{tab:para}}
\tablewidth{0pt}
\tablehead{ Target Name &  $N^{\rm{Gal,*}}_{\rm{H}} $ & $N_{\rm{H}} $ & $N_{\rm{XIS}} $ & 
$\Gamma$  &  $f_{\rm{scat}} $ & $R_{\rm torus}$ & $A_{\rm K\alpha}$ & $kT $ & $kT_2 $  
& $\chi^2/d.o.f.$  
\\ \rule{0pt}{3ex}
(1) & (2) & (3) & (4) & (5) & (6) & (7) & (8) & (9) & (10) & (11)
} 
\startdata 
NGC~2655 & $2.18$ & $26.1^{+2.7}_{-1.7}$ & $1.05^{+0.94}_{-0.23}$ & $1.77^{+0.19}_{-0.07}$ & 
$3.5^{+3.4}_{-1.4}$ & $0.06 (< 1.71) $ & $5.3\pm1.2$ & $0.18^{+0.06}_{-0.05}$ 
& $0.74\pm 0.04$ & $186.6/167$ \\
NGC~3718 & $1.06$ & $1.29^{+0.12}_{-0.11}$ & $0.68^{+0.31}_{-0.15}$ & $1.86^{+0.13}_{-0.09}$ 
& $1.5^{+1.0}_{-0.7}$  & $0.37 (< 1.57) $ & $< 0.7$ & - & - & $214.9/206$ & \\
NGC~3998 & $1.01$ & $0.008^{+0.006}_{-0.005}$  & $0.88^{+0.16}_{-0.12}$ & $1.84\pm0.01$ 
& - & $ < 0.10$  & $< 0.7$ & - & - & $950.8/982$ & \\ 
NGC~4102 & $1.68$ & $81^{+34}_{-27}$ & $0.25^{+0.14}_{-0.12}$ & $1.84^{+0.05}_{-0.13}$ & 
$8.3^{+3.6}_{-2.3}$ & $ < 0.14$ & $14.7\pm2.7$ & $0.61^{+0.11}_{-0.18}$ & 
$1.10^{+0.18}_{-0.13}$ & $299.3/265$ \\ 
NGC~4138 & $1.25$ & $7.82^{+0.72}_{-0.68}$ & $0.42^{+0.13}_{-0.09}$ & $1.59^{+0.17}_{-0.06}$ &
 $0.82^{+0.52}_{-0.39}$ & $0.06 (< 0.81)$ & $6.1\pm1.5$ & $0.33^{+0.19}_{-0.06}$ & - 
& $78.1/94$ & \\
NGC~4258 &  $1.60$ & $12.10^{+0.52}_{-0.51}$ & $1.07^{+0.37}_{-0.15}$ 
& $1.77^{+0.10}_{-0.06}$ & $16.2^{+5.6}_{-3.3}$ & $0.07 (< 0.86)$ & $2.5^{+1.2}_{-1.1}$ 
& $0.34^{+0.02}_{-0.03}$ & $0.86^{+0.05}_{-0.07}$ & $640.8/573$ \\
NGC~4395 &  $1.85$ & $1.55^{+0.28}_{-0.23}$ & $1.55^{+0.28}_{-0.23}$ & $1.49^{+0.15}_{-0.10}$ 
& $40^{+11}_{-9}$ & $1.67^{+1.73}_{-0.92}$ & $4.0\pm1.2$ & $0.19\pm0.03$ & - 
& $332.7/354$ \\
NGC~4941 &  $2.17$ & $70^{+17}_{-14}$ & $0.54^{+0.33}_{-0.18}$ & $1.82^{+0.25}_{-0.26}$ & 
$0.94^{+1.59}_{-0.60}$ & $0.42^{+0.47}_{-0.26}$ & $6.2\pm1.1$ & $0.87^{+0.14}_{-0.16}$  & - 
& $33.1/43$ \\ 
NGC~5273 & $0.916$ & $2.60^{+0.12}_{-0.11} $ & $4.10^{+1.36}_{-0.85}$ & 
$1.57^{+0.07}_{-0.06}$ & $6.6^{+2.6}_{-1.7}$ & $0.44^{+0.45}_{-0.35}$  & $9.7\pm1.7$ 
& - & - & $586.4/563$ &  \\ 
NGC~5643 & $8.01$ &  $94^{+61}_{-32}$ & $1.18^{+0.81}_{-0.48}$  & $1.57^{+0.37}_{-0.31}$ & 
$ 10.1^{+23.2}_{-7.4}  $ & $0.86 (< 3.52)$ & $19.2\pm2.1$  
& $0.18\pm 0.04$ & $0.88\pm 0.14$ & $109.5/114$ 
\enddata
\tablecomments{Columns: 
(1) Galaxy name. 
(2) Galactic absorption in units of $10^{20}$ cm$^{-2}$. 
(3) Intrinsic absorption in units of $10^{22}$ cm$^{-2}$. 
(4) Normalization ratio of the primary X-ray emission in the 
{\it Suzaku}/FI-XISs spectrum to the {\it Swift}/BAT one. 
(5) Photon index of the power-law component.
(6) Scattering fraction in units of $\%$. 
(7) Relative reflection strength ($R = \Omega/2\pi$) of the {\tt pexrav} 
model. 
(8) Photon flux of the {\tt zgauss} model in units of $10^{-6}$ 
${\rm photons}$ ${\rm cm}^{-2} $ ${\rm s}^{-1}$.  
(9)-(10) Temperatures of the {\tt apec} models. 
(11) Reduced chi-squared over degrees of freedom. 
}
\tablenotetext{*}{The parameter is fixed.}
\end{deluxetable*}

\tabletypesize{\footnotesize}
\begin{deluxetable*}{cccccccccccccccccccccccc}
\tablecaption{Fluxes and Luminosities\label{tab:flux}}
\tablewidth{0pt}
\tablehead{ Target Name & 
$F^{\rm BI-XIS}_{0.5-2} $ & $F^{\rm FI-XISs}_{2-10} $  &  $F^{\rm PIN,*}_{10-50} $ & $F^{\rm BAT}_{10-50} $ & 
$L^{\rm BI-XIS}_{0.5-2} $ & $L^{\rm FI-XISs}_{2-10} $  & $L^{\rm PIN,*}_{10-50} $  & $L^{\rm BAT}_{10-50} $  & 
$\log \lambda_{\rm Edd}$  \\ \rule{0pt}{3ex}
(1) & (2) & (3) & (4) & (5) & (6) & (7) & (8) & (9) & (10) 
} 
\startdata 
NGC~2655 & 
$3.1\times10^{-13}$ & $2.0\times10^{-12}$ & -- & $6.8\times10^{-12}$ & 
$2.4\times10^{41}$ & $3.9\times10^{41}$ & -- & $5.1\times10^{41}$ & 
$-3.2$ \\ 
NGC~3718 & 
$4.5\times10^{-13}$ & $3.3\times10^{-12}$ & $5.4\times10^{-12}$ & $7.3\times10^{-12}$ & 
$9.4\times10^{40}$ & $1.3\times10^{41}$ & $1.9\times10^{41}$ & $2.5\times10^{41}$ & 
$-3.8$ \\  
NGC~3998 & 
$4.1\times10^{-12}$ & $6.6\times10^{-12}$ & $8.0\times10^{-12}$ & $9.1\times10^{-12}$ & 
$2.0\times10^{41}$ & $3.0\times10^{41}$ & $3.6\times10^{41}$ & $4.1\times10^{41}$ & 
$-4.5$ \\ 
NGC~4102 & 
$1.2\times10^{-12}$ & $1.5\times10^{-12}$ & $4.0\times10^{-12}$ & $1.3\times10^{-11}$ & 
$1.5\times10^{41}$ & $1.8\times10^{41}$ & $2.0\times10^{41}$ & $6.5\times10^{41}$ & 
$-3.3 $ \\
NGC~4138 & 
$7.3\times10^{-14}$ & $2.2\times10^{-12}$ & -- & $1.4\times10^{-11}$ & 
$5.0\times10^{40}$ & $1.0\times10^{41}$ & --  & $4.4\times10^{41}$ & 
$-2.9 $ \\ 
NGC~4258 & 
$2.5\times10^{-12}$ & $5.2\times10^{-12}$ & $1.2\times10^{-11}$ & $1.2\times10^{-11}$ & 
$5.1\times10^{40}$ & $6.3\times10^{40}$ & $8.7\times10^{40}$ & $8.2\times10^{40}$ & 
$-3.9 $ \\ 
NGC~4395 & 
$6.9\times10^{-13}$ & $4.7\times10^{-12}$ & $1.8\times10^{-11}$ & $1.2\times10^{-11}$ & 
$3.9\times10^{39}$ & $9.4\times10^{39}$ & $3.3\times10^{40}$ & $2.3\times10^{40}$ & 
$-2.7 $ \\
NGC~4941 & 
$8.4\times10^{-14}$ & $8.3\times10^{-13}$ & $6.7\times10^{-12}$ & $1.0\times10^{-11}$ & 
$1.4\times10^{41}$ & $2.1\times10^{41}$ & $3.4\times10^{41}$ & $5.5\times10^{41}$ & 
$-2.4 $ \\ 
NGC~5273 & 
$3.1\times10^{-13}$ & $8.1\times10^{-12}$ & $2.2\times10^{-11}$ & $5.6\times10^{-12}$ & 
$8.4\times10^{40}$ & $2.0\times10^{41}$ & $4.5\times10^{41}$ & $1.1\times10^{41}$ & 
$ -2.5 $ \\
NGC~5643 & 
$4.4\times10^{-13}$ & $1.3\times10^{-12}$ & $9.0\times10^{-12}$ & $7.7\times10^{-12}$ & 
$6.6\times10^{40}$ & $1.3\times10^{41}$ & $3.3\times10^{41}$ & $2.8\times10^{41}$ & 
$ -2.3 $ 
\enddata
\tablecomments{
(1) Galaxy name. 
(2)--(5) Observed fluxes in the 0.5--2 keV (BI-XIS), 2--10 keV (FI-XISs), 
10--50 keV (PIN), and 10--50 keV (BAT) bands. 
(6)--(9) Absorption-corrected luminosities in the 0.5--2 keV (BI-XIS), 2--10 
keV (FI-XISs), 10--50 keV (PIN), and 10--50 keV (BAT) bands. 
(10) Eddington ratio.
}
\tablenotetext{*}{
According to the XIS or HXD nominal positions, the fluxes and luminosities 
are divided by 1.16 or 1.18 to take into account the instrumental 
cross-calibration factor between FI-XISs and HXD/PIN. 
} 
\end{deluxetable*}

\subsection{Broad Iron-K$\alpha$ Line}\label{subsec:broad} 

In this section, we investigate possible contribution of a relativistically 
broadened iron-K$\alpha$ line in our spectra. We exclude NGC 4102, NGC 4941, 
and NGC 5643 from this study, which show heavy absorptions 
($N_{\rm H} > 5\times10^{23}$ cm$^{-1}$) preventing detailed studies of the 
broad iron-K$\alpha$ line feature. 

First, we fit the broadband X-ray spectra with the best-fit model obtained in 
Section~\ref{sec:broad_ana_2} plus a disk-reflection component 
({\tt rdblur*pexmon}). The {\tt rdblur} model calculates the relativistic 
blurring around a Schwarzschild black hole \citep{Fab89}, and the 
{\tt pexmon} model reproduces the same reflection continuum as {\tt pexrav} 
together with emission lines of Fe K$\alpha$, Fe K$\beta$, and Ni K$\alpha$ 
computed in a self-consistent way \citep{Nan07}. This disk component is 
corrected for the same absorption and time variability as those for the 
primary X-ray emission. Because the two reflection components from the disk 
and distant matter are strongly coupled in a spectral fit, we fix the disk 
reflection strength $R_{\rm disk}$. For this purpose, we refer to the study by 
\cite{Geo91}, who calculated the predicted equivalent width of the 
iron-K$\alpha$ line from an annulus slab irradiated by a point source with a 
power-law spectrum. We thus determine $R_{\rm disk}$ in the {\tt pexmon} model 
that reproduces the predicted equivalent width of the iron-K$\alpha$ line, by
assuming an inclination angle ($\theta_{\rm inc}$), a photon index, and a ratio
 between the scale height of the X-ray source and the inner radius of the 
annulus slab ($r_{\rm in}$). Here $\theta_{\rm inc}$ is set to 30$^\circ$ and 
60$^\circ$ for type-1 ($N_{\rm H} < 10^{22}$ cm$^{-2}$) and type-2 
($N_{\rm H} > 10^{22}$ cm$^{-2}$) AGNs, respectively.  For NGC 4258, we adopt 
$\theta_{\rm inc} = 80^\circ$, which was precisely estimated by Maser 
observations \citep{Herr99}. The photon index is tied to that of the primary 
X-ray component. We adopt the scale height of $10 r_{\rm g}$ ($r_{\rm g}$ is 
the gravitational radius), a typical value measured in some AGNs (e.g., 
\citealt{Mor08}; \citealt{Mor12}). Then, we consider two cases, 
$r_{\rm in} = 10 r_{\rm g}$ and $100 r_{\rm g}$, which correspond to 
$R_{\rm disk} \simeq 0.6$ and $R_{\rm disk} \simeq 0.1$, respectively. In the 
{\tt rdblur} model, the outer radius and radial emissivity index are set to 
$r_{\rm out} = 10^5 r_{\rm g}$ and $\beta = -3$, respectively. 

We find that adding the disk reflection component with the above parameters 
does not significantly improve the fit in all 7 objects ($\Delta \chi^2 < 3$).
 Thus, our data do not require a cold standard disk extending down to 
10--100 $r_{\rm g}$, although its presence cannot be strictly ruled out within
 the quality of our data (see below). 

To derive upper limits of the flux of a broad iron-K$\alpha$ line, we replace 
{\tt pexmon*rdblur} with {\tt diskline+pexrav*rdblur} and fit the XIS spectra 
in the 3--9 keV (3--8 keV for BI-XIS) band. The parameters 
($\theta_{\rm inc}$, $\beta$, $r_{\rm in}$, and $r_{\rm out}$) of the {\tt diskline} model are set 
to the same as in the {\tt rdblur} model. We leave the normalizations of the 
{\tt zgauss}, {\tt diskline}, and {\tt zpowerlw} models as free parameters.
Table~\ref{tab:broad_line} summarizes the line fluxes and equivalent widths 
(EW$^{\rm obs}_{\rm disk}$) of the broad iron-K$\alpha$ line in the case of 
$r_{\rm in} = 10 r_{\rm g}$, which gives more conservative upper limits than 
the case of $r_{\rm in} = 100 r_{\rm g}$.  To check whether the assumed 
equivalent widths based on \cite{Geo91} are consistent with the observed 
upper limits, we also calculate the corrected equivalent width 
(EW$^{\rm cor}_{\rm disk}$) with respect to the continuum composed of the 
primary power-law and disk-reflection components. We find that in NGC 3998 
the upper limit of EW$^{\rm cor}_{\rm disk} < 20$ eV is inconsistent with 
the assumed one ($\approx$ 120 eV for $r_{\rm in} = 10 r_{\rm g}$ and 
 $\approx$ 20--30 eV for $r_{\rm in} = 100 r_{\rm g}$), implying that the inner 
disk radius may be much larger than $100 r_{\rm g}$. In the other objects, the
 assumed equivalent widths are consistent within the observed upper limits, 
and hence the possible presence of a cold standard disk with $r_{\rm in} =$ 
10--100 $r_{\rm g}$ is not ruled out.

\begin{deluxetable}{cllllll}
\tabletypesize{\footnotesize}
\tablecaption{Flux and Equivalent Widths of Broad Iron-K$\alpha$ Line\label{tab:broad_line}}
\tablewidth{0pt}
\tablehead{ 
Target Name & $N_{\rm disk} $ & $ {\rm EW}_{\rm disk}^{\rm obs}$ & $ {\rm EW}_{\rm disk}^{\rm cor}$ \\ \rule{0pt}{3ex}
(1) & (2) & (3) & (4)
}
\startdata
NGC~2655 &  $< 12.2$ & $< 250$ & $< 300$ \\
NGC~3718 &  $< 4.4$ & $< 90$ & $< 90$ \\
NGC~3998 &  $< 0.9$ & $< 20$ & $< 20$ \\ 
NGC~4138 &  $< 17.7$ & $< 240$ & $< 250$ \\
NGC~4258 &  $< 9.3$ & $< 100$ & $< 130$ \\ 
NGC~4395 &  $< 5.6$ & $< 140$ & $< 200$ \\
NGC~5273 &  $< 2.1$ & $< 80$ & $< 80$ 
\enddata
\tablecomments{
(1) Galaxy name. 
(2) Photon flux of the {\tt diskline} model in units of $10^{-6}$ 
${\rm photons}$ ${\rm cm}^{-2} $ ${\rm s}^{-1}$. 
(3) Observed equivalent width of the {\tt diskline} model in units of eV.
(4) Corrected equivalent width of the {\tt diskline} model in units of eV. 
}
\end{deluxetable}

\subsection{Narrow Iron-K$\alpha$ Line}\label{subsec:narrow} 

\begin{figure*}[!ht]
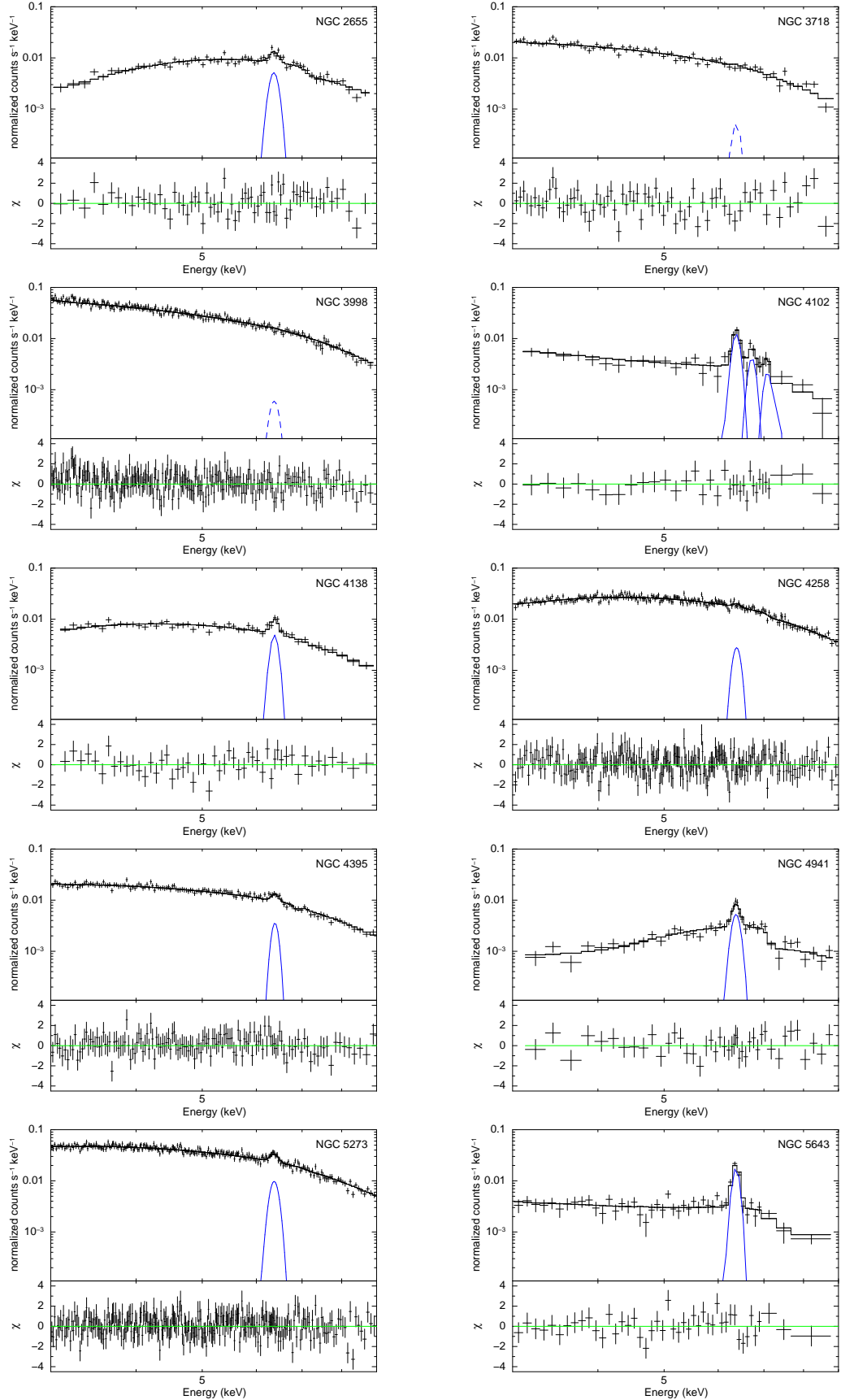

\centering
\includegraphics[scale=0.25,angle=-90]{ngc_2655_iron_line.ps} \hspace{1cm}
\includegraphics[scale=0.25,angle=-90]{ngc_3718_iron_line.ps}
\includegraphics[scale=0.25,angle=-90]{ngc_3998_iron_line.ps} \hspace{1cm}
\includegraphics[scale=0.25,angle=-90]{ngc_4102_iron_line.ps}
\includegraphics[scale=0.25,angle=-90]{ngc_4138_iron_line.ps} \hspace{1cm}
\includegraphics[scale=0.25,angle=-90]{ngc_4258_iron_line.ps}
\includegraphics[scale=0.25,angle=-90]{ngc_4395_iron_line.ps} \hspace{1cm}
\includegraphics[scale=0.25,angle=-90]{ngc_4941_iron_line.ps} 
\includegraphics[scale=0.25,angle=-90]{ngc_5273_iron_line.ps} \hspace{1cm}
\includegraphics[scale=0.25,angle=-90]{ngc_5643_iron_line.ps}
\caption{
Spectra of FI-XISs in the 3--9 keV band. The black crosses and solid line 
represent the data and best-fit model, respectively. Emission lines that are 
detected significantly are shown with the blue solid lines, while those not 
significantly detected are shown with the blue dashed lines. The lower panel 
plots the residuals. 
}
\label{fig:narrow_line}
\end{figure*}

Similarly, we obtain the flux and observed equivalent width of the narrow 
iron-K$\alpha$ line (EW$^{\rm obs}_{\rm gauss}$) by fitting the XIS narrow band 
(3--9 keV) spectra with the best-fit models obtained in 
Section~\ref{sec:broad_ana_2}. Only the normalizations of the {\tt zgauss} and
 {\tt zpowerlw} components are allowed to vary in this analysis. 
Figure~\ref{fig:narrow_line} plots the spectra with the best-fit model. For 
easier comparison with model predictions (see Section~\ref{sec:ikeda}), we 
also calculate the corrected equivalent width (EW$^{\rm cor}_{\rm gauss}$) with 
respect to the continuum composed of the primary power-law and 
torus-reflection components. The results are summarized in 
Table~\ref{tab:narrow_line}.

\begin{deluxetable}{cccccccc}
\tabletypesize{\footnotesize}
\tablecaption{Flux and Equivalent Widths of Narrow Iron-K$\alpha$ Line\label{tab:narrow_line}}
\tablewidth{0pt}
\tablehead{ Target Name & $N_{\rm gauss}$ & $ {\rm EW}_{\rm gauss}^{\rm obs}$ & $ {\rm EW}_{\rm gauss}^{\rm cor}$ &  \\ \rule{0pt}{3ex}
(1) & (2) & (3) & (4)
}
\startdata
NGC~2655 & $5.3\pm1.2$ &  $150 \pm30$ & $160\pm40$ \\ 
NGC~3718 & $< 0.7$  &  $< 20$ & $< 20$ \\ 
NGC~3998 & $< 0.7$  &  $< 10$ & $< 10$ \\ 
NGC~4102 & $14.7^{+2.8}_{-2.7}$ & $920 \pm490$ & $560\pm300$ \\   
NGC~4138 & $6.0\pm1.5$ &  $190 \pm50$ & $80\pm20$ \\ 
NGC~4258 & $2.5\pm1.2$ &  $30 \pm20$ & $40\pm20$ \\ 
NGC~4395 & $3.9\pm1.2$ &  $70 \pm20$ & $80\pm20$ \\ 
NGC~4941 & $6.2\pm1.1$ & $380 \pm70$ & $240\pm50$ \\ 
NGC~5273 & $9.6\pm1.7$    & $90 \pm20$ & $100\pm20$ \\ 
NGC~5643 & $19.0\pm2.1$  &  $1160 \pm330$ & $1970\pm560$ 
\enddata
\tablecomments{
(1) Galaxy name. 
(2) Photon flux of the {\tt zgauss} model in units of $10^{-6}$ 
${\rm photons}$ ${\rm cm}^{-2} $ ${\rm s}^{-1}$. 
(3) Observed equivalent width of the {\tt zgauss} model in units of eV. 
(4) Corrected equivalent width of the {\tt zgauss} model in units of eV.
}
\end{deluxetable}

\section{DISCUSSIONS}\label{sec:dis_con}

\subsection{Summary of X-ray Spectral Analysis}

We have systematically analyzed the broadband X-ray spectra in the 0.5--200 
keV band of ten nearby LLAGNs observed with {\it Suzaku} and {\it Swift}/BAT. 
The spectra are basically well reproduced with the base-line model composed 
of absorbed primary X-ray emission, a scattered component, and a 
reflection component from distant matter with a fluorescence iron-K$\alpha$ 
line. Some spectra require additional components, such as optically-thin 
thermal emission and emission/absorption lines. The spectra of NGC 3718 and 
NGC 3998 show no narrow iron-K$\alpha$ line, whereas it is significantly 
detected in the other objects. In the following discussions, we refer to the 
X-ray luminosities measured with {\it Suzaku} and {\it Swift}/BAT for the 
three low mass LLAGNs (NGC 4395, NGC 5273, and NGC 5643) and the other LLAGNs,
 respectively.

\subsection{State of Accretion Disk}

\subsubsection{Broad Iron-K Line}

In Section~\ref{subsec:broad}, we show that a relativistically broadened 
iron-K$\alpha$ line from a standard disk extending down to 10--100 $r_{\rm g}$ is not 
required from the spectral fit, even though we cannot, in most of the targets,
 rule out its presence within the statistical errors. Only upper limits of the
 line flux are derived. In the case of NGC 3998, the tight constraint on the 
equivalent width of the broad iron-K$\alpha$ line ($< 20$ eV) indicates that the 
standard disk must be truncated at a radius larger than $100 r_{\rm g}$. This 
is consistent with the result by \cite{Nem14}, who reported that a RIAF model
 fit to the SED of NGC 3998 requires a truncation radius larger than 
$1000 r_{\rm g}$. 

\subsubsection{Correlation between Photon Index and Eddington Ratio}\label{subsubsec:cor_edd_ind}

\begin{figure}[!ht]
\centering
\includegraphics[scale=0.52,angle=-90]{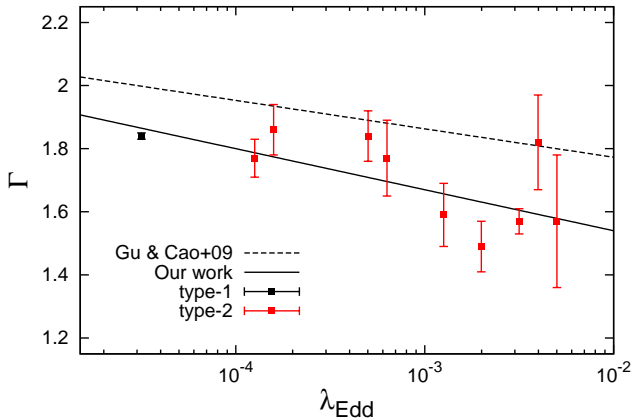}
\caption{                                     
Photon index plotted against Eddington ratio. The solid line shows the 
best-fit correlation based on our sample, while the dashed line was derived 
by \cite{Gu09}. Errors correspond to 1$\sigma$.
}
\label{fig:index_vs_edd-ratio}
\end{figure}

The broadband coverage of our data including the hard X-ray band above 10 keV 
has enabled us to most reliably determine the photon index of the intrinsic 
power-law continuum even for obscured AGNs. Figure~\ref{fig:index_vs_edd-ratio} 
plots the best-fit photon index against the Eddington ratio for our sample. 
We obtain a significant negative correlation with a form of 
$\Gamma = (-0.13\pm0.02)\log(\lambda_{\rm Edd})+(1.28\pm0.07)$ by a $\chi^2$ 
fit (errors are 1$\sigma$). The slope is consistent with that found by 
\cite{Gu09}, who derived a form of 
$\Gamma = (-0.09\pm0.03)\log(\lambda_{\rm Edd})+(1.55\pm0.07)$ for low 
Eddington-ratio ($\log \lambda_{\rm Edd} \lesssim -2$) AGNs. 
\cite{You11} also derived a negative but steeper slope of $-0.31\pm0.06$ for 
lower Eddington-ratio AGNs ($\log \lambda_{\rm Edd} < -3$). To compare our result 
with that of \cite{Gu09} in Figure~\ref{fig:index_vs_edd-ratio}, we convert the 
2--10 keV luminosity into the bolometric one with 
a 2--10 keV  bolometric correction factor of 10 (\citealt{Ho09}; \citealt{Vas09}).

It is known that the slope between the photon index and Eddington ratio turns 
into a positive value at high Eddington-ratio AGNs. For instance, \cite{She08} 
 reported a positive slope of $0.31\pm0.01$ for AGNs with 
$\log \lambda_{\rm Edd} \gtrsim -2$. A similar result was also obtained by 
\cite{Bri13}. As suggested by some authors (e.g., \citealt{She08}; 
\citealt{Gu09}), the $\Gamma$-$\lambda_{\rm Edd}$ correlations for high and 
low Eddington-ratio AGNs imply that the state of the accretion disk 
changes according to the Eddington ratio. 
The positive correlation at the high Eddington-ratio regime can be
explained in terms of standard disk as follows. 
A hot corona located above the disk produces Comptonized emission of seed photons provided from 
the disk. Hence, as the disk luminosity (or accretion rate) increases, electrons in 
the corona are more effectively cooled through inverse Compton scattering, 
leading to a smaller Compton y-parameter, and hence a softer spectrum. 
There are some numerical studies on the 
$\Gamma$-$\lambda_{\rm Edd}$ correlation for low Eddington-ratio AGNs under a RIAF assumption. 
For instance, the Convection dominated accretion flow model developed by 
\cite{Bal01} predicts a negative correlation,
whereas the Advection dominated accretion flow
model by \cite{Nara98} predicts a positive one, 
which is inconsistent with the observations.
Thus, this relation can be used to test RIAF models at low mass-accretion rates.

\subsection{Torus Structure}

\subsubsection{Application of Torus Model}\label{sec:ikeda}

The flux (or equivalent width) of a narrow iron-K$\alpha$ line at 6.4 keV is
useful to constrain the geometry and column density of circumnuclear matter, 
which we call the ``torus''. 
To constrain the torus opening angle with respect to the polar axis, we utilize 
the Ikeda torus model \citep{Ike09}, which is a Monte-Carlo based 
numerical spectral model. The torus structure has a nearly spherical shape 
with two conical-shaped holes along the rotation axis, and is defined by three
 parameters: the hydrogen column density at the equatorial plane 
($N_{\rm H}^{\rm eq}$), the half opening angle ($\theta_{\rm oa}$), and the 
inclination angle ($\theta_{\rm inc}$). Hence, 
$\theta_{\rm inc} < \theta_{\rm oa}$ for type-1 AGNs and 
$\theta_{\rm inc} > \theta_{\rm oa}$ for type-2 AGNs. As done in previous works 
(\citealt{Kawa13}; \citealt{Taz13}), we fix $N_{\rm H}^{\rm eq}$ at the 
observed line-of-sight column density ($N_{\rm H}$) and a photon index of the 
incident X-ray spectrum according to the best-fit value listed in 
Table~\ref{tab:para}. 

\begin{figure*}[!ht]
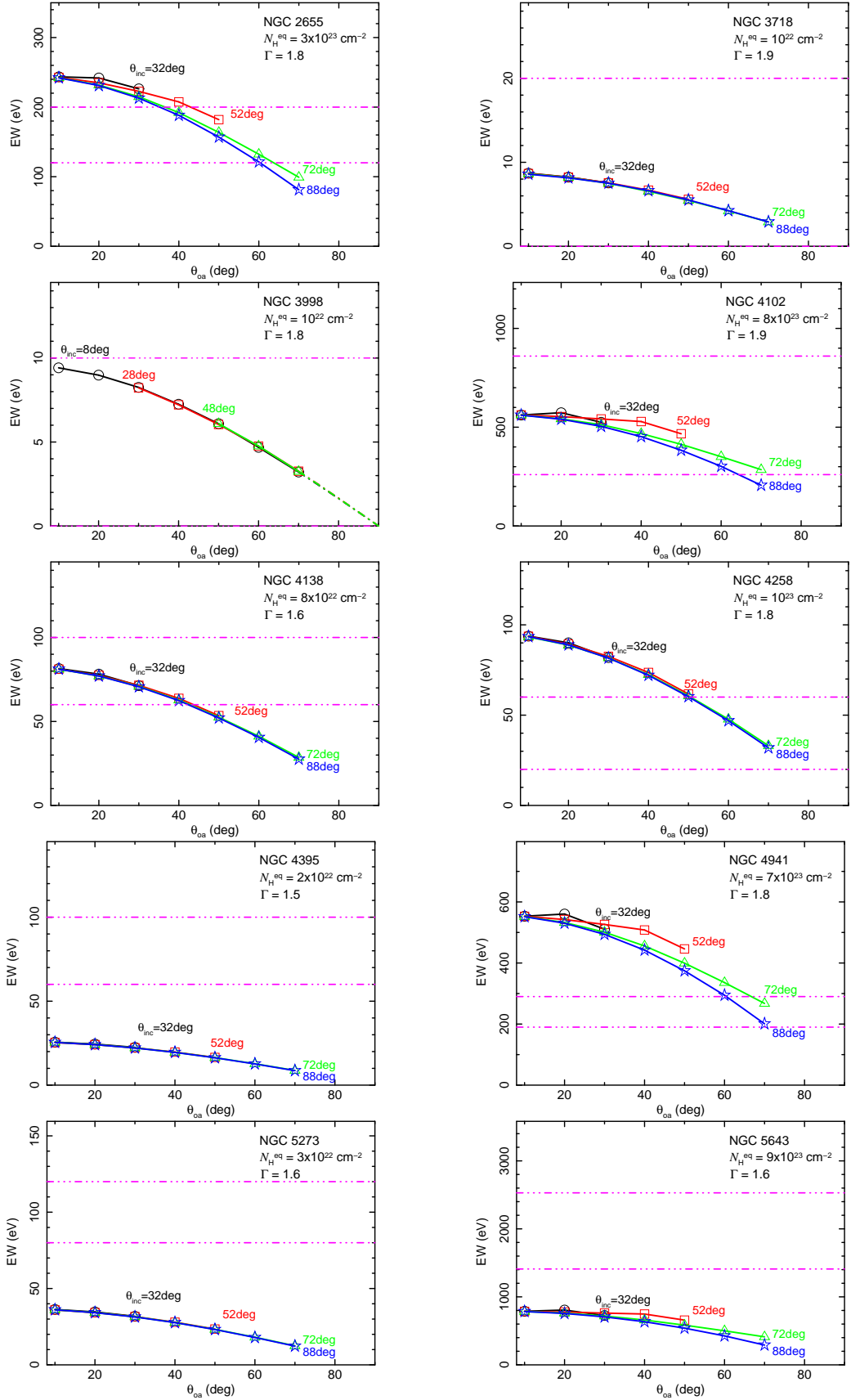

\centering
\includegraphics[scale=0.25,angle=-90]{Abs30_Gamma1.8.ps} \hspace{1cm}
\includegraphics[scale=0.25,angle=-90]{Abs1_Gamma1.9.ps} 
\includegraphics[scale=0.25,angle=-90]{Abs1_Gamma1.8.ps}  \hspace{1cm}
\includegraphics[scale=0.25,angle=-90]{Abs80_Gamma1.9.ps} 
\includegraphics[scale=0.25,angle=-90]{Abs8_Gamma1.6.ps} \hspace{1cm}
\includegraphics[scale=0.25,angle=-90]{Abs10_Gamma1.8.ps}  
\includegraphics[scale=0.25,angle=-90]{Abs2_Gamma1.5.ps} \hspace{1cm}
\includegraphics[scale=0.25,angle=-90]{Abs70_Gamma1.8.ps}  
\includegraphics[scale=0.25,angle=-90]{Abs3_Gamma1.6.ps} \hspace{1cm}
\includegraphics[scale=0.25,angle=-90]{Abs90_Gamma1.6.ps} 
\caption{
Predicted equivalent widths of the iron-K$\alpha$ line as a function of the 
torus half-opening angle with different inclinations, calculated from the 
torus model by \cite{Ike09}. The 90\% confidence upper and lower limits of 
EW$^{\rm cor}_{\rm gauss}$ are represented with the two horizontal lines 
(magenta).
}
\label{fig:ikeda_model}
\end{figure*}

Figure~\ref{fig:ikeda_model} shows the predicted equivalent width of the 
iron-K$\alpha$ line as a function of $\theta_{\rm oa}$ for several different 
inclinations. Because simulated spectra for $\theta_{\rm oa} >70^\circ$ are not
 available in the Ikeda torus model, for type-1 AGNs, we simply extrapolate 
the result at $\theta_{\rm oa}=70^\circ$ towards larger opening angles by 
assuming that the equivalent width is proportional to the volume of the torus.
The observed range of EW$^{\rm cor}_{\rm gauss}$ at the 90\% confidence limits is 
represented with the dot-dot-dot-dashed horizontal lines (magenta). 

As noticed from Figure~\ref{fig:ikeda_model}, the half opening angles of NGC 
3718 and NGC 3998 cannot be constrained if we assume 
$N_{\rm H}^{\rm eq} = N_{\rm H}$, although their weak iron-K$\alpha$ line fluxes
 and low line-of-sight absorptions suggest that their tori are little 
developed, having a small covering fraction and/or a small averaged column 
density. By contrast, NGC 2655, NGC 4102, NGC 4138, NGC 4258, and NGC 4941 
have moderately developed tori, fulfilling $N_{\rm H} > 7\times10^{22}$ cm$^{-2}$ and/or 
$\theta_{\rm oa} < 70^\circ$. For NGC 4395, NGC 5273, and NGC 5643, 
the Ikeda model largely underpredicts the equivalent widths under the 
assumption of $N_{\rm H}^{\rm eq} = N_{\rm H}$. We find that their equatorial 
hydrogen column densities must be larger than $6\times10^{22}$ cm$^{-2}$, 
$8\times10^{22}$ cm$^{-2}$, and $1.4\times10^{24}$ cm$^{-2}$, respectively, to 
consistently explain the observed values of EW$^{\rm cor}_{\rm gauss}$.

To summarize, we reveal that there are at least two types of LLAGNs in terms 
of torus structure: those with moderately developed tori 
($\Omega/2\pi > 0.34$ corresponding to $\theta_{\rm oa} < 70^\circ$, or 
$N^{\rm eq}_{\rm H} > 5\times10^{22}$ cm$^{-2}$) or those with little developed  
tori. 
The 
variety in the torus structure of LLAGNs is a new result obtainable
only by detailed spectral analysis of individual objects.
Previous works based on
the type-2 AGN fraction have given only the ``averaged'' torus covering
factor, which decreases towards lower luminosities (\citealt{Bur11}; \citealt{Bri11b}).

\subsubsection{Luminosity Dependence} 

\cite{Ricci14} pointed out that, in order to discuss the torus structure of 
type-2 AGNs, the luminosity ratio between the iron-K$\alpha$ line and 10--50 
keV continuum is a better indicator of the torus covering factor than the 
iron-K$\alpha$ equivalent width 
because the continuum flux above 10 keV is less affected by absorption than 
that at 6.4 keV. Using an AGN sample observed with {\it Suzaku}, they found 
that the \ironc\ ratio decreases with luminosity in both type-1 and type-2 
AGNs at $L_{\rm 10-50\hspace{1mm} keV} \gtrsim 10^{42}$ erg s$^{-1}$. These results can be 
explained if the torus covering fraction becomes smaller at higher 
luminosities, and are consistent with the anti-correlation between the 
absorbed-AGN fraction and luminosity in this luminosity range (e.g., \citealt{Ued03}).

\begin{figure}[!ht]
\centering
\includegraphics[scale=0.52,angle=-90]{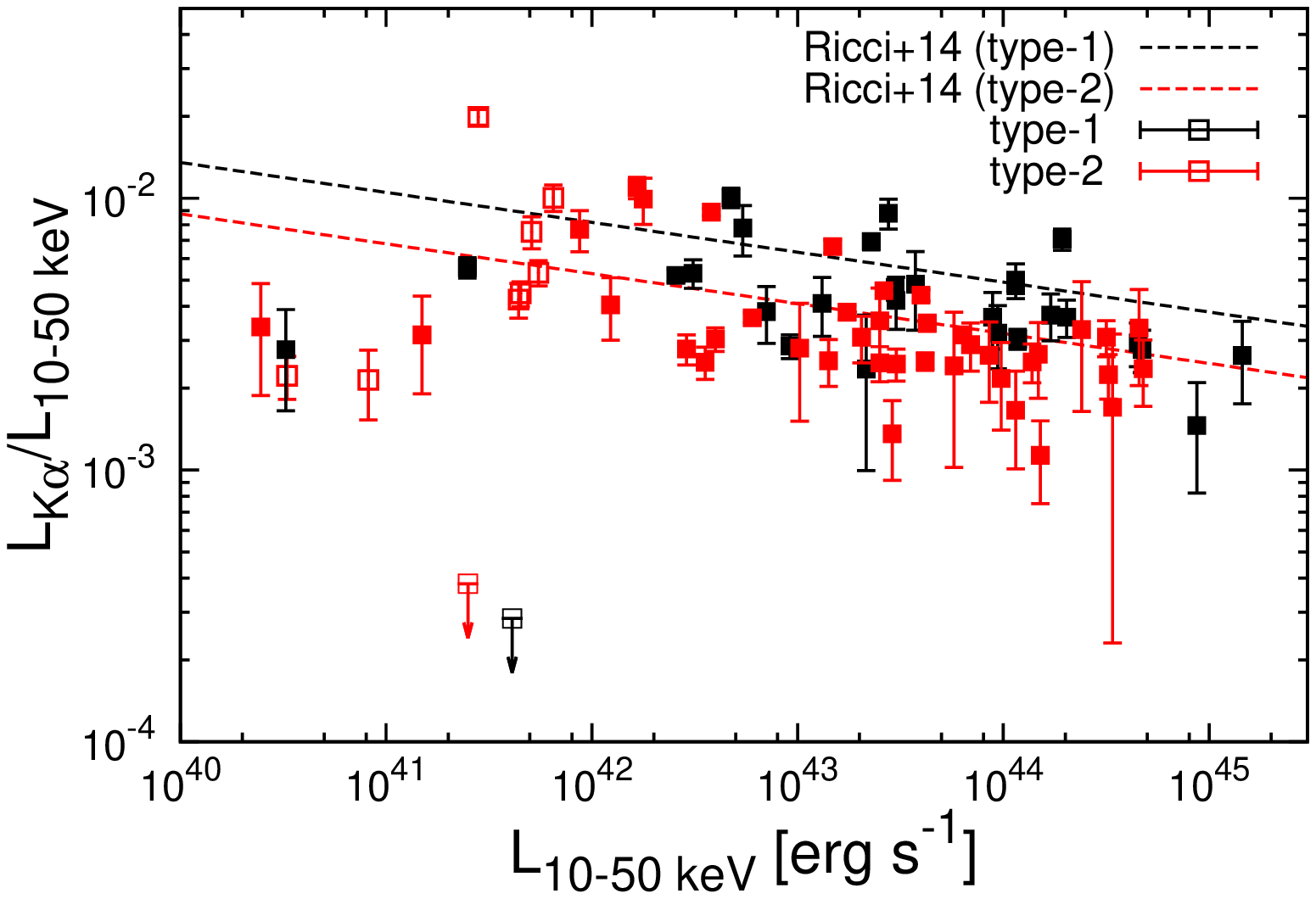}
\caption{
The luminosity ratio between the iron-K$\alpha$
line and 10--50 keV continuum plotted
against the luminosity in the 10--50 keV band.
The open squares are the data of our sample, while the filled ones are 
taken from \cite{Ricci14}.
The arrows represent the upper limits for NGC 3718 and NGC 3998.
The attached errors or upper limits are 1$\sigma$. 
}
\label{fig:line_vs_lum}
\end{figure}

We plot \ironc\ against luminosity of our LLAGN sample together with the 
original sample of \cite{Ricci14} in Figure~\ref{fig:line_vs_lum}. Here we 
exclude Compton-thick AGNs ($N_{\rm H} > 10^{24}$ cm$^{-2}$) because the heavy 
obscuration also strongly affects the iron-K$\alpha$ line flux and the 
relation may become complex \citep[see][]{Ricci14}. The dashed lines in the 
figure are the best-fit linear regression forms for type-1 and type-2 AGNs 
obtained by \cite{Ricci14}. 

Figure~\ref{fig:line_vs_lum} shows a trend that the \ironc\ ratio has a peak 
around $L_{\rm 10-50\hspace{0.5mm}keV} = 10^{42}$ erg s$^{-1}$, from which the 
average value rapidly declines towards lower luminosities. This agrees with the
 implication from hard X-ray ($>10$ keV) surveys that the absorbed AGN 
fraction peaks at $\sim 10^{42-43}$ erg s$^{-1}$ and decreases with decreasing 
luminosity (\citealt{Bec09}, \citealt{Bur11}). Thus, our study using the 
narrow iron-K$\alpha$ line supports the picture that the averaged solid angle 
of AGN tori is not a monotonically decreasing function with luminosity, and 
that irradiation is not the only key factor that determines the torus 
structure. We note that there is a large scatter of \ironc\ in the LLAGNs, 
as discussed in the previous subsection.

\subsubsection{Eddington Ratio Dependence}\label{sec:lumEdd2iron}

\begin{figure}[!ht]
\centering
\includegraphics[scale=0.52,angle=-90]{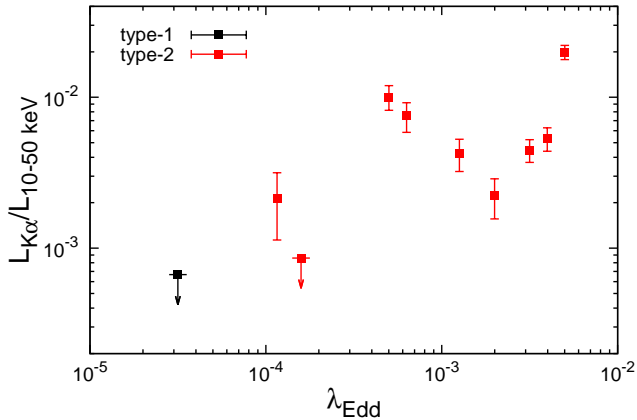}
\caption{
The luminosity ratio between the iron-K$\alpha$ line and 10--50 keV luminosity
 plotted against Eddington ratio for our LLAGN sample. The arrows represent 
the upper limits at the 90\% confidence level for NGC 3718 and NGC 3998. 
}
\label{fig:ew_vs_edd-rat}
\end{figure}

Figure~\ref{fig:ew_vs_edd-rat} shows the Eddington ratio dependence of 
\ironc\ based on our LLAGN sample. A positive correlation is seen between the
 \ironc\ ratio and $\lambda_{\rm Edd}$. If we exclude NGC 4258 (the second left
 point), whose narrow iron-K$\alpha$ line may not be originated from the torus
 \citep{Rey09}, there is a trend that all LLAGNs with 
$\lambda_{\rm Edd} \lesssim 2\times10^{-4}$ have \ironc\ smaller than $10^{-3}$.
 We thus infer that the Eddington ratio is a key parameter that affects the 
torus structure of LLAGNs: tori are hardly developed below a critical value of
 $\lambda_{\rm Edd}$.  Because the sample size is limited, however, 
we would need a larger number of LLAGNs to draw a robust conclusion.
The lack of the torus may lead to the shortage of mass 
supply through it, hence reducing the mass-accretion rate. In an opposite way,
 the torus is not formed because of the lack of powerful outflow from the 
inner accretion disk at low mass-accretion rates. Thus, to understand the 
reason of the \ironc\ versus $\lambda_{\rm Edd}$ correlation, we need to reveal
 the physical origin(s) of AGN tori.

\subsubsection{Correlation between Luminosities in X-ray and MIR Bands}  

Many studies found strong luminosity correlations between the X-ray and
mid-infrared (MIR) bands (e.g., \citealt{Lut04}; \citealt{Ram07}; \citealt{Gan09}; 
\citealt{Asm11}; \citealt{Mas12}; \citealt{Ichi12}; \citealt{Asm15}; \citealt{Mat15}; 
\citealt{Ste15}). Such correlation is 
expected because the dust torus heated by an AGN emits blackbody radiation in 
the MIR band. \cite{Asm11} find no Eddington-ratio dependence of the 
X-ray/MIR (12 $\mu$m) luminosity ratio using a sample that covers a wide 
Eddington ratio range of $\log \lambda_{\rm Edd} = -6 - 0$.

\begin{figure}[thd]
\centering
\includegraphics[scale=0.52,angle=-90]{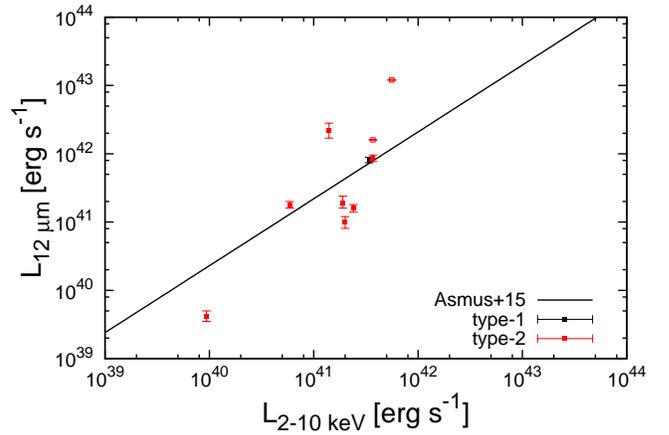}
\caption{
Luminosity correlation of our LLAGNs between the 12 $\mu$m and 2--10 keV 
bands.  The solid line shows the regression line derived by \cite{Asm15}. The 
12 $\mu$m luminosities from the {\it WISE} catalog are represented with 
unfilled squares, while those taken from \cite{Asm15} with filled squares. 
Errors correspond to 1$\sigma$.
} 
\label{fig:x-ray_vs_ir}
\end{figure}

Figure~\ref{fig:x-ray_vs_ir} shows the $L_{\rm X}$ versus $L_{\rm MIR}$ correlation
obtained from our LLAGN sample. The regression line derived by \cite{Asm15} 
from a large local AGN sample is plotted by the solid line. \cite{Asm15} took
 only the nuclear MIR emission by separating from that in the host galaxy, 
using subarcsecond resolution imaging data. We refer to the photometric data 
in the 12 $\mu$m band of \cite{Asm15} for our objects whenever available, and 
to the Wide-Field Infrared Survey Explorer ({\it WISE}) catalog \citep{Wri10}
 for the rest. Due to the poor angular resolution of {\it WISE} (6.5 arcsec), 
however, its flux may be contaminated from that of the host galaxy and hence 
should be regarded as an upper limit of the nuclear emission.

We find that our LLAGNs also approximately follow the same correlation as 
found by \cite{Asm15} for higher luminosity AGNs, regardless of AGN types and 
Eddington ratios. This is expected if a (moderate size of) torus is 
present in LLAGNs. There is a possibility that synchrotron emission from the 
jets or blackbody radiation from the truncated disk may contribute to the MIR 
luminosity in some LLAGNs (e.g., \citealt{Ho08}; \citealt{Nem14}). Our result 
in Section~\ref{sec:lumEdd2iron} suggests that the low Eddington ratio LLAGNs 
($\lambda_{\rm Edd} \lesssim 2\times10^{-4}$) do not have developed tori. In 
these objects, the MIR luminosity would be indeed dominated by these emission 
mechanisms other than thermal emission from AGN-heated dust.

\subsubsection{MIR Selection of LLAGNs}  

\begin{figure}[h]
\centering
\includegraphics[scale=0.51,angle=-90]{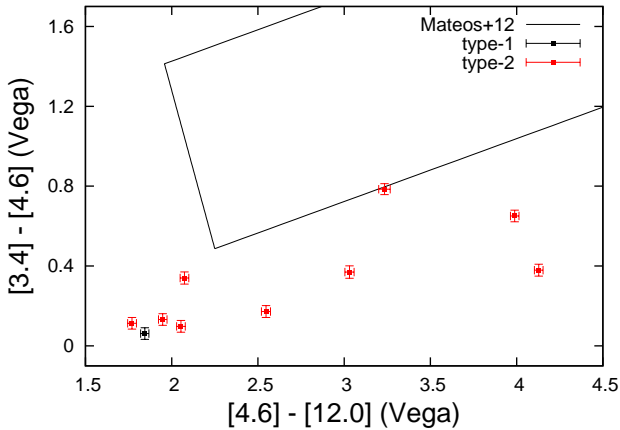}
\caption{
MIR color-color plot of our LLAGN sample, where the horizontal and vertical 
axes represent the 4.6--12.0 $\mu$m and  3.4--4.6 $\mu$m colors, respectively.
The AGN wedge defined by \cite{Mate12} is surrounded by the solid lines. 
Errors correspond to 1$\sigma$.
}
\label{fig:agn_wedge}
\end{figure}

MIR SEDs are often used to identify AGNs by detecting a power-law component 
originating in AGN-heated dust (e.g., \citealt{Pol07}). The spectral index in 
the MIR band of luminous AGNs are typically $\alpha \lesssim -0.5$ in the 
form of $ f_\nu = \nu^{\alpha}$ (e.g., \citealt{Alo06}). Accordingly, 
\cite{Mate12} defined the so-called AGN-wedge region in the MIR color-color 
diagram based on the three {\it WISE} bands (3.4 $\mu$m, 4.6 $\mu$m, and 12.0 
$\mu$m) in order to efficiently select luminous AGNs. 
Figure~\ref{fig:agn_wedge} shows the color-color plot of our LLAGN sample based on the {\it WISE} 
data together with the AGN-wedge region. As noticed, all objects are not located 
within this region. This is attributable to significant contamination from 
the host galaxy in the {\it WISE} MIR fluxes. This is an example demonstrating
 difficulty in identifying AGNs using MIR data alone.

\section{CONCLUSIONS}\label{sec:con} 

We have performed a systematic broadband (0.5--200 keV) X-ray spectral 
analysis of ten local LLAGNs (with intrinsic 14--195 keV luminosities of 
$<10^{42}$ erg s$^{-1}$) observed with {\it Suzaku} and {\it Swift}/BAT. The 
main conclusions are summarized as follows. 

\begin{enumerate}

\item The broadband X-ray spectra can be basically reproduced with an
absorbed cut-off power law often accompanied by a reflection 
component from distant cold matter and a narrow fluorescence iron-K$\alpha$  
line, and optically-thin thermal emission from the host galaxy.

\item In all objects, a relativistically blurred reflection component from a 
standard disk is not required from the spectra. We find the negative correlation 
between the photon index and Eddington ratio. These results are consistent with a  
picture that an optically-thin RIAF is formed at innermost radii in LLAGNs.

\item Applying a torus model by \cite{Ike09}, we find eight objects showing a 
significant narrow iron-K$\alpha$ emission line have a moderately developed 
torus with an equatorial column density of $\log N^{\rm eq}_{\rm H} > 22.7$ or 
a half opening-angle of $\theta_{\rm oa} < 70^\circ$. The tori of the two 
LLAGNs without a narrow iron-K$\alpha$ line are likely little developed.

\item The luminosity ratio between the iron-K$\alpha$ line and 10--50 keV 
continuum (\ironc ) has a peak around $L_{\rm 10-50\hspace{0.5mm}keV} = 10^{42}$ 
erg s$^{-1}$, from which the average value rapidly declines towards lower 
luminosities. This result indicates that AGN irradiation is not the only key 
factor that determines the torus structure. Low Eddington-ratio LLAGNs with
$\lambda_{\rm Edd} \lesssim 2\times10^{-4}$ show small \ironc\ ratios. We infer
 that the Eddington ratio is a key parameter that affects the torus structure 
of LLAGNs. 

\item Regardless of Eddington ratio, our LLAGNs follow the same luminosity 
correlation between the hard X-ray and MIR bands as found for more luminous 
AGNs. This implies that other emission mechanisms than AGN-heated dust are 
responsible for the MIR emission in low Eddington-ratio LLAGNs without 
developed tori. 

\end{enumerate}

\acknowledgments

We thank the referee for the comments, which helped us to improve the 
quality of the manuscript. We are grateful to C. Ricci for kindly 
providing us with the data. Part of this work was financially supported 
by the Grant-in-Aid for JSPS 
Fellows for young researchers (T.K) and for Scientific Research 26400228 (YU).
This research has made use of the NASA/ IPAC Infrared Science Archive, which 
is operated by the Jet Propulsion Laboratory, California Institute of 
Technology, under contract with the National Aeronautics and Space 
Administration.

\appendix 
\def\thesection{\Alph{section}}

\section{{\it XMM-Newton} Observations of the ULX in NGC 5643}\label{subsec:xmm} 

To model the spectrum of the ULX in NGC 5643, we use the {\it
XMM-Newton} data (OBSID = 0601420101) obtained in 2009, because the exposure 
($\sim 50$ ksec) is longer than that of the other {\it XMM-Newton}
observation ($\sim 10$ ksec) in 2003. The Standard Analysis Software
(version 13.5.0) is utilized to reduce the data of the EPIC cameras
(MOS-1, MOS-2, and pn).
For all cameras, the source events are extracted from a circular region
of a 26'' radius, and the background events are taken from a source-free 
circular region with the same radius. The spectra of MOS-1 and MOS-2 are
summed. 

We find that a cut-off power law plus a disk-blackbody component well
reproduces the spectra in the 0.3--7 keV band ($\chi^2/d.o.f$ =
183.4/154). Here the cut-off energy is fixed at 6 keV, a value observed
from a typical ULX, NGC~1313 X-1 \citep{Miz07}; the fit is not
significantly improved by leaving the cut-off energy as a free parameter
($\chi^2/d.o.f$ = 182.4/153). The best-fit parameters are 
a photon index of $\Gamma = 0.8$, a disk-blackbody normalization of 3.7 $\times10^{36}$ erg s$^{-1}$
kpc$^{-2}$, and a disk temperature of 0.35 keV. When we analyze the {\it
Suzaku} spectra of NGC~5643, we always include this best-fit ULX model.

\end{document}